# Effects of Divalent Cations on Diffusion Dynamics of Biological Water Confined between Lipid Membranes

Minho Lee[1,2], Jinwon Park[1,2], Ji-Hyun Kim[1,2], Minhaeng Cho[3,4]*, and Jaeyoung Sung[1,2]*

[1] Global Science Research Center for Systems Chemistry (GCSC), Seoul, 06974, Republic of Korea

[2]Department of Chemistry, Chung-Ang University, Seoul, 06974, Republic of Korea

[3]Center for Molecular Spectroscopy and Dynamics, Institute for Basic Science (IBS), Seoul 02841, Republic of Korea

[4]Department of Chemistry, Korea University, Seoul 02841, Republic of Korea

Corresponding authors: *mcho@korea.ac.kr (Minhaeng Cho), *jaeyoung@cau.ac.kr (Jaeyoung Sung)

# Abstract

Biological water is an ionic solution containing both monovalent and divalent ions. However, the effects of divalent ions on the dynamics of biological water remain largely unknown. Here, we investigate how the transport dynamics of water molecules nanoconfined between lipid membranes depends on the concentration of calcium ($Ca^{2+}$) and magnesium ($Mg^{2+}$) ions by using molecular dynamics simulations and the generalized transport equation for biological water. We find that the diffusion coefficient of biological water monotonically increases with $Ca^{2+}$ ion concentration but exhibits a largely opposite, non-monotonic dependence on $Mg^{2+}$ concentration. The deviation of the water molecules' displacement distribution from the Gaussian also shows distinct dependence on the concentrations of $Mg^{2+}$ and $Ca^{2+}$. These contrasting behaviors originate from the different hydration radii of these divalent ions and their distinct effects on the interfacial structure and dynamics of biological water. The relaxation of the lateral displacement distribution of water molecules toward a Gaussian is determined by the time-correlation function of diffusion coefficient fluctuations, whose relaxation time increases with salt concentrations. The primary source of the lateral diffusion coefficient fluctuation is thermal motion of water molecules in the longitudinal direction, along which microscopic environments surrounding a water molecule, including the functional groups of lipid membrane and ion concentrations, drastically change.

# I. Introduction

Divalent ions in biological water play essential roles in the functions of living cells. Calcium ions ($Ca^{2+}$) regulate membrane fusion[1,2], cell action potentials[3], intracellular signal transmission[4,5], and muscle contraction[6], while also acting as a critical cofactor in blood coagulation[7]. Magnesium ions ($Mg^{2+}$) modulate the activity of cellular enzymes[8] and the interaction between F-actin and the cell membrane[9,10], and are indispensable for the structural stabilization of nucleic acids[11,12] as well as the bioactivity of ATP molecules[13]. Performing these biological functions, divalent cations interact with membrane components and alter the structural organization of lipid bilayers. Both experimental and theoretical studies have revealed that $Ca^{2+}$ and $Mg^{2+}$ ions induce distinct structural alterations in phosphatidylcholine (PC) lipid membranes. For example, $Ca^{2+}$ induces condensation of PC lipid membranes, whereas $Mg^{2+}$ does not[14]. This difference reflects ion-dependent hydration and membrane binding characteristics: $Ca^{2+}$ has a small hydration radius and binds tightly to lipid headgroups, bridging adjacent headgroups and promoting membrane condensation, whereas $Mg^{2+}$ retains a larger hydration shell and only weakly binds to lipid molecules[15–20]. These divalent cations also have distinct effects on the structure of interfacial water molecules near membranes. $Ca^{2+}$ disrupts the dipole alignment of interfacial water molecules by weakening their interactions with the membrane surface, whereas $Mg^{2+}$ induces only minor perturbations[15,21,22]. Such structural perturbations are closely related to the dynamical behavior of interfacial water molecules, which mediate a wide range of biological processes[23,24].

Extensive research effort has been directed to investigate the effects of divalent cations on the structures of lipid bilayer interfacial water[14–22]; however, their influence on the dynamics of the interfacial water remains largely unexplored. A notable exception is Seto and Yamada's quasi-elastic neutron scattering (QENS) experiment, which measures the populations of various hydration water species with different diffusion coefficients[25]. Their results showed opposite effects of $Ca^{2+}$ and $Mg^{2+}$ on the relative populations of hydration water species; the fraction of bound water with a smaller diffusion coefficient decreases in 0.45 M $CaCl_2$ solution but increases in 0.45 M $MgCl_2$ solution compared to nonionic biological water, suggesting ion-specific modulation of interfacial water dynamics. However, microscopic mechanism underlying this observation remains elusive. For nonionic intermembrane water, the previous molecular dynamics (MD) simulation studies have reproduced experimental observables such as the water molecules' diffusion coefficient [26–28] and the electron density profiles across lipid bilayers[29]. These MD simulation studies provide atomistic insights into microscopic structures and dynamics of intermembrane water molecules[15,17,30–34]. However, for ionic biological water, such MD simulation studies have mostly focused only on the structure of interfacial water and the physical understanding of distinct $Ca^{2+}$ and $Mg^{2+}$ effects on the transport dynamics of biological water has yet to be attained.

In this work, we employ all-atom MD simulations and the generalized transport equation to investigate the transport dynamics of water molecules in divalent ionic solutions nanoconfined between PC lipid bilayers, specifically dimyristoylphosphatidylcholine (DMPC) bilayers, while systematically controlling the concentrations of $CaCl_2$ and $MgCl_2$ in the solutions. DMPC bilayers were chosen because the PC lipids are one of the major components of cellular membranes[35], which has been widely investigated as a representative membrane model system[16,25,36–40]. Our MD simulation results show that the intermembrane water molecules exhibit a qualitatively different dependence of the displacement distribution on the concentration of $Ca^{2+}$ and $Mg^{2+}$. For biological water confined between membranes, the mean square displacement (MSD) and mean diffusion coefficient monotonically increase with $Ca^{2+}$ ion concentration but exhibit largely opposite, non-monotonic dependence on $Mg^{2+}$ concentration. The transient deviation of the water molecules' displacement distribution from a Gaussian also shows distinct dependence on the concentrations of $Mg^{2+}$ and $Ca^{2+}$. We find that these contrasting behaviors originate from different hydration radii of these divalent ions and their distinct effects on the interfacial structure and dynamics of biological water. By analyzing the transient super-Gaussian displacement distribution of water molecules using the recently proposed transport equation of biological water[41,42], we extract the variance and the time-correlation function of the diffusion coefficient fluctuations in our biological water system. We find that the relative variance of the diffusion coefficient increases with $Mg^{2+}$ ion concentration but tends to decrease with $Ca^{2+}$ ion concentration. On the other hand, the relaxation time of the water displacement distribution to a Gaussian increases with ionic concentration for both $Mg^{2+}$ and $Ca^{2+}$ solutions.

This paper is organized as follows. In Section II, we present the theoretical framework underlying our description of the transport dynamics of water nanoconfined between lipid bilayers. Section III describes the details of the MD simulation systems and computational methods. The results and discussion are presented in section IV, and the main results are summarized in section V with concluding remarks.

# II. Theory

The transport equation governing the lateral transport dynamics of ionic biological water confined between lipid membranes is given by[41]:

$$\hat{\dot{p}}(\boldsymbol{r}_\parallel, \boldsymbol{\Gamma}, s) = \widehat{\mathcal{D}}_\parallel(\boldsymbol{\Gamma}, s)\boldsymbol{\nabla}_\parallel^2 \hat{p}(\boldsymbol{r}_\parallel, \boldsymbol{\Gamma}, s) + L(\boldsymbol{\Gamma})\hat{p}(\boldsymbol{r}_\parallel, \boldsymbol{\Gamma}, s) \tag{1}$$

Here, $\hat{p}(\boldsymbol{r}_\parallel, \boldsymbol{\Gamma}, s)$ is the Laplace transform of the joint probability density, $p(\boldsymbol{r}_\parallel, \boldsymbol{\Gamma}, t)$, where the center of mass (COM) of a water molecule is located at lateral position $\boldsymbol{r}_\parallel$ $[= (x, y)]$ and the hidden state is at $\boldsymbol{\Gamma}$ at time $t$. The hidden state variable $\boldsymbol{\Gamma}$ represents the entire set of dynamical variables that influence the lateral dynamics of a water molecule's COM, including the water molecule's rotational degrees of freedom, the local configuration and concentrations of the surrounding ions, water, and lipid molecules. In Eq. (1), $\boldsymbol{\nabla}_\parallel^2$, $\widehat{\mathcal{D}}_\parallel(\boldsymbol{\Gamma}, s)$, and $L(\boldsymbol{\Gamma})$ denote, respectively, the Laplacian of the COM coordinates, $\boldsymbol{r}_\parallel$, in the two-dimensional plane parallel to the lipid membrane, Laplace transform of the diffusion kernel of a water molecule under environmental state $\boldsymbol{\Gamma}$ [41], and a mathematical operator describing the time-evolution of hidden state variable $\boldsymbol{\Gamma}$. Throughout, $\hat{f}(s)$ and $\hat{\dot{f}}(s)$ denote the Laplace transform of $f(t)$ and $\dot{f}(t)[\equiv \partial_t f(t)]$, i.e., $\hat{f}(s) = \int_0^\infty dt\, e^{-st} f(t)$ and $\hat{\dot{f}}(s) = \int_0^\infty dt\, e^{-st} \partial_t f(t)$.

The first two nonvanishing moments of the lateral displacement distribution, $\Delta_2(t)$ and $\Delta_4(t)$, can be obtained from Eq. (1) as[41]:

$$\hat{\Delta}_2(s) = \frac{4}{s^2}\left\langle \widehat{\mathcal{D}}_\parallel(s) \right\rangle \tag{2}$$

$$\hat{\Delta}_4(s) = 4s\hat{\Delta}_2(s)^2\left[1 + s\hat{C}_{\mathcal{D}_\parallel}(s)\right] \tag{3}$$

where $\Delta_n(t)$ denotes the n-th moment of the time-dependent distribution of the water displacement, $\Delta\boldsymbol{r}_\parallel(t)[\equiv \boldsymbol{r}_\parallel(t) - \boldsymbol{r}_\parallel(0)]$, in the lateral direction. The bracket notation, $\langle \dots \rangle$, represents an average over the equilibrium distribution, $P_{eq}(\boldsymbol{\Gamma})$, of hidden state $\boldsymbol{\Gamma}$. The MSD, or $\Delta_2(t)$, depends only on the mean diffusion kernel, $\langle \mathcal{D}_{\parallel}(t) \rangle$, which equals to the velocity autocorrelation function (VAF) scaled by 2 in our system[41]. On the other hand, the fourth moment, $\Delta_4(t)$, depends not only on $\langle \mathcal{D}_{\parallel}(t) \rangle$, but also on the diffusion kernel correlation (DKC), $C_{\mathcal{D}_{\parallel}}(t)$, defined by

$$\hat{C}_{\mathcal{D}_{\parallel}}(s) = \int d\mathbf{\Gamma} \int d\mathbf{\Gamma}_0 \frac{\delta \hat{\mathcal{D}}_{\parallel}(\mathbf{\Gamma}, s)}{\langle \hat{\mathcal{D}}_{\parallel}(s) \rangle} \hat{\mathcal{G}}(\mathbf{\Gamma}, s | \mathbf{\Gamma}_0) \frac{\delta \hat{\mathcal{D}}_{\parallel}(\mathbf{\Gamma}_0, s)}{\langle \hat{\mathcal{D}}_{\parallel}(s) \rangle} P_{eq}(\mathbf{\Gamma}_0) \tag{4}$$

where $\delta \hat{\mathcal{D}}_{\parallel}(\mathbf{\Gamma}, s)$ and $\hat{\mathcal{G}}(\mathbf{\Gamma}, s | \mathbf{\Gamma}_0)$ denote, respectively, $\hat{\mathcal{D}}_{\parallel}(\mathbf{\Gamma}, s) - \langle \hat{\mathcal{D}}_{\parallel}(s) \rangle$ and the Green's function, i.e., the conditional probability that the surrounding environment is at state $\mathbf{\Gamma}$ at time $t$, given that it was initially at state $\mathbf{\Gamma}_0$. The Green's function obeys $\partial_t \mathcal{G}(\mathbf{\Gamma}, t | \mathbf{\Gamma}_0) = L(\mathbf{\Gamma}) \mathcal{G}(\mathbf{\Gamma}, t | \mathbf{\Gamma}_0)$ with the initial condition, $\mathcal{G}(\mathbf{\Gamma}, 0 | \mathbf{\Gamma}_0) = \delta(\mathbf{\Gamma} - \mathbf{\Gamma}_0)$. Equations (2) and (3) enable us to obtain the non-Gaussian parameter (NGP), $\alpha_2(t) [\equiv \Delta_4(t) / (2\Delta_2(t)^2) - 1]$, which measures the deviation of the lateral displacement distribution from a Gaussian[43,44], in terms of the VAF and DKC. That is to say, we can extract the time profiles of the VAF and DKC from the MSD and NGP time profiles using Eqs. (2) and (3)[41,42].

At long times where the VAF has negligible values, thermal motion of water molecules exhibits diffusive dynamics. The MSD and NGP of the lateral diffusion of water molecules have the following approximate expressions[42]

$$\Delta_2(t) \cong 4 \langle D_{\parallel} \rangle t, \tag{5}$$

$$\alpha_2(t) \cong \frac{2}{t^2} \int_0^t d\tau \ (t - \tau) \eta_{D_{\parallel}}^2 \phi_{D_{\parallel}}(\tau), \tag{6}$$

where $\langle D_{\parallel} \rangle$, $\phi_{D_{\parallel}}(\tau)$, and $\eta_{D_{\parallel}}^2$ designate, respectively the mean diffusion coefficient $\langle D_{\parallel} \rangle = \int_0^{\infty} d\tau \, \langle \mathcal{D}_{\parallel}(\tau) \rangle$, and the normalized time correlation function, $\langle \delta D_{\parallel}(t) \delta D_{\parallel}(0) \rangle / \langle \delta D_{\parallel}^2 \rangle$, and the relative variance of the lateral diffusion coefficient $\langle \delta D_{\parallel}^2 \rangle / \langle D_{\parallel}^2 \rangle$. DKC defined in Eq. (4) reduces to $\eta_{D_{\parallel}}^2 \phi_{D_{\parallel}}(t)$ at long times where the MSD linearly increases with time. Noting that Eq. (6) is equivalent to $t^2 \alpha_2(t) \cong 2 \int_0^t d\tau_2 \int_0^{\tau_2} d\tau_1 \, \eta_{D_{\parallel}}^2 \phi_{D_{\parallel}}(\tau_1)$, we can extract the accurate time profile of $\eta_{D_{\parallel}}^2 \phi_{D_{\parallel}}(t)$ from the NGP time profile, i.e.,

$$\eta_{D_\parallel}^2 \phi_{D_\parallel}(t) \cong 2^{-1}\, d^2\big(t^2 \alpha_2(t)\big)/dt^2 \tag{7}$$

Alternatively, an approximate estimation of $\phi_{D_\parallel}(t)$ and $\eta_{D_\parallel}^2$ can be made based on the following equations[42]:

$$\phi_{D_\parallel}(t) = \frac{\langle \delta D_\parallel(t) \delta D_\parallel(0) \rangle}{\langle \delta D_\parallel^2 \rangle} \cong \langle \delta D_\parallel^2 \rangle^{-1} \int_0^{\ell} dz \int_0^{\ell} dz_0\, \delta D_\parallel(z) G(z,t|z_0) \delta D_\parallel(z_0) p_{eq}(z_0), \tag{8}$$

$$\eta_{D_\parallel}^2 = \langle \delta D_\parallel^2 \rangle / \langle D_\parallel \rangle^2 = \left[ \langle D_\parallel^2 \rangle - \langle D_\parallel \rangle^2 \right] / \langle D_\parallel \rangle^2, \tag{9}$$

with $\langle D_\parallel^m \rangle = \int_0^{\ell} dz\, D_\parallel^m(z) p_{eq}(z)$ where $p_{eq}(z)$ is the equilibrium probability distribution of water molecules along the $z$-axis, defined as $p_{eq}(z) = \rho_{\mathrm{H_2O}}(z) / \int_0^{\ell} dz'\, \rho_{\mathrm{H_2O}}(z')$. The essential assumption underlying Eq. (8) is that the most important hidden variable that affects the lateral thermal motion of an interfacial water molecule is the distance, $z$, between the water molecule's COM and the center of the membrane. This is a legitimate assumption because microscopic environments interacting with a water molecule, including the functional groups of lipid molecules, ion concentrations, and hydrogen bond network, drastically change with $z$[36,42,45,46]. In Eqs. (8) and (9), $\ell$, $D_\parallel(z)$, and $G(z,t|z_0)$ denote, respectively, the distance between the center positions of the two membranes confining water molecules (see Fig. 1), the $z$-dependent lateral diffusion coefficient, and Green's function governing the thermal motion of water molecules in the direction perpendicular to the membrane, with the initial condition, $G(z,0|z_0) = \delta(z - z_0)$.

Using our theory, we have analyzed the MSD and NGP time profiles of water displacement distribution obtained from the MD simulation, for various concentrations of divalent ions in our biological water system. From the quantitative analyses, we can extract the VAF and DKC of water molecules for each ion species and concentration. We note here that the long-time profile of the DKC, directly extracted from the MSD and NGP profiles, is in excellent agreement with $\eta_{D_\parallel}^2 \phi_{D_\parallel}(t)$ calculated from Eq. (7). We also find that the time profile of $\eta_{D_\parallel}^2 \phi_{D_\parallel}(t)$ calculated from Eqs. (8) and (9) also serve as a good approximation of the long-time DKC.

# III. Computational details

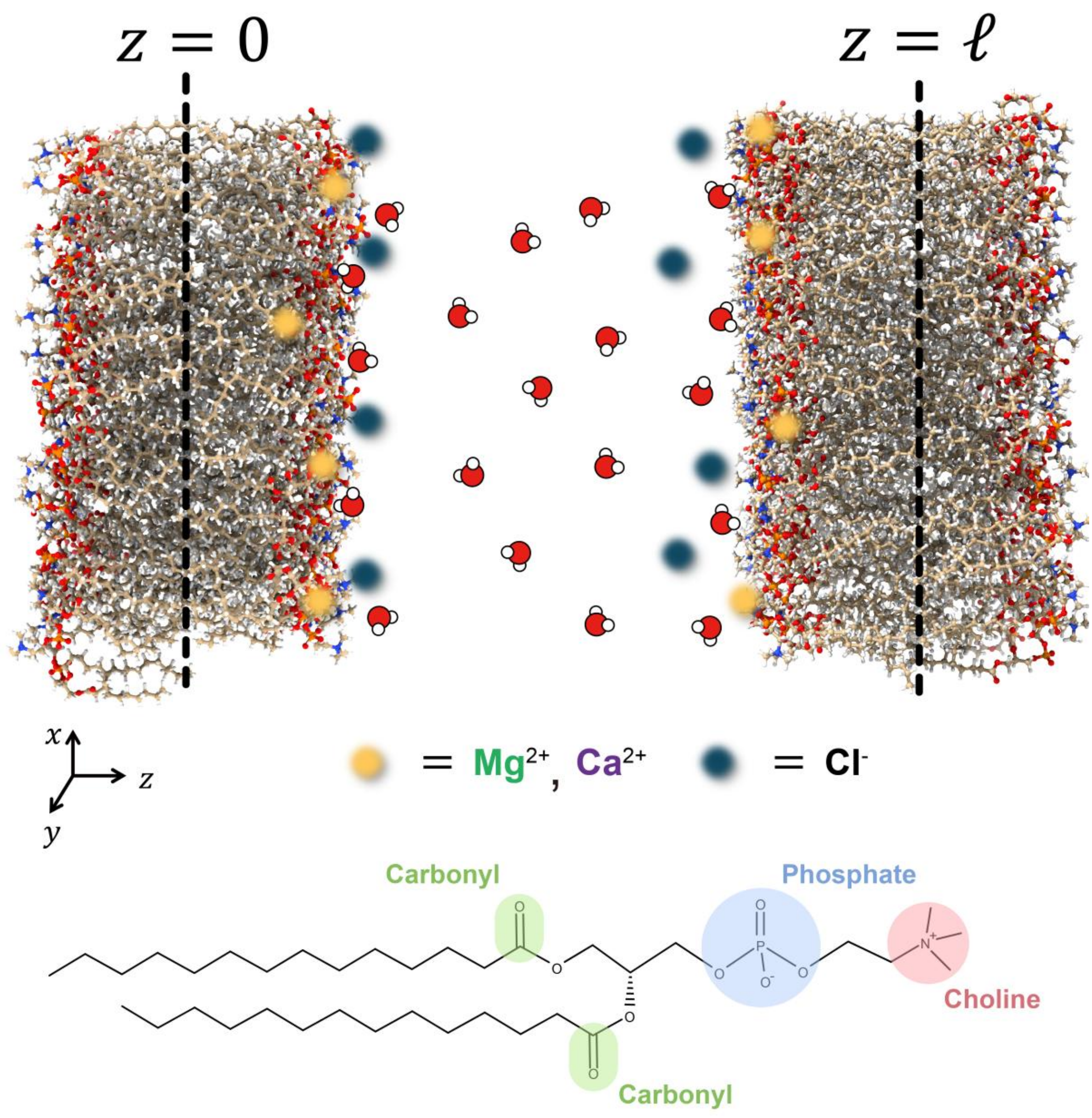

**FIG 1.** Schematic representation of our MD simulation system: $MgCl_2$ and $CaCl_2$ solutions nanoconfined between two DMPC lipid bilayers. The simulations were performed at three different salt concentrations (0.1M, 0.4M, and 0.8M) for each ionic solution as well as for the salt-free water nanoconfined between the lipid bilayers.

We conducted simulations for seven different systems, each containing 128 DMPC lipid molecules and salt solutions with varying ion types and concentrations (Fig. 1). The water-to-lipid ratio was set to 37:1, identical to that used in a previous experimental study[25]. The ion density was systematically varied by adding either $CaCl_2$ or $MgCl_2$ at concentrations of 0 M (salt-free), 0.1M, 0.4M, and 0.8M. The similar $CaCl_2$ and $MgCl_2$ concentration values were used in previous studies of biological water systems[14,22,25]. Initial structures of our system were generated using CHARMM-GUI membrane builder[47–50]. The force-field parameters for DMPC, water molecules, and ions were taken from the AMBER Lipid 21[51], SPC/E, and Li-Merz 12-6-4 parameter sets[52–55], respectively. MD simulations were performed using the AMBER 21 program package under periodic boundary conditions to account for the confinement of water molecules between the bilayers. Long-range electrostatic interactions were treated using the particle mesh Ewald (PME) method[56], while a cutoff distance of 10 Å was applied to the Lennard-Jones interactions and the real-space part of the Ewald sum. The time step of our simulations was set to 1 fs.

Prior to the production run, an equilibration procedure was performed. The system underwent a 10,000-step energy minimization from the initial configuration using both the steepest descent and the conjugate gradient methods. Then, the system was rapidly heated from 0 K to 100 K over 40 ps, using the Langevin thermostat with collision frequency of 1.0 $ps^{-1}$ and weak restraints on the lipid molecules with a force constant of 10 kcal $mol^{-1}$ $Å^{-2}$. Subsequently, a gradual heating from 100 K to 318 K over 2 ns was performed with the same thermostat setting and restraints. After the heating process, *NpT* simulations were carried out at 318 K, using the Langevin thermostat with anisotropic pressure scaling (1 atm) without restraining lipids. For the salt-free case, a 50 ns *NpT* simulation was performed, whereas 1-2 μs *NpT* simulations were conducted for the other systems to ensure equilibration, confirmed by the stabilization of the area per lipid during the relaxation process (see Fig. 2). We also examined whether a similar area per lipid relaxation process occurs in more complex membrane environments, as described in Appendix F.

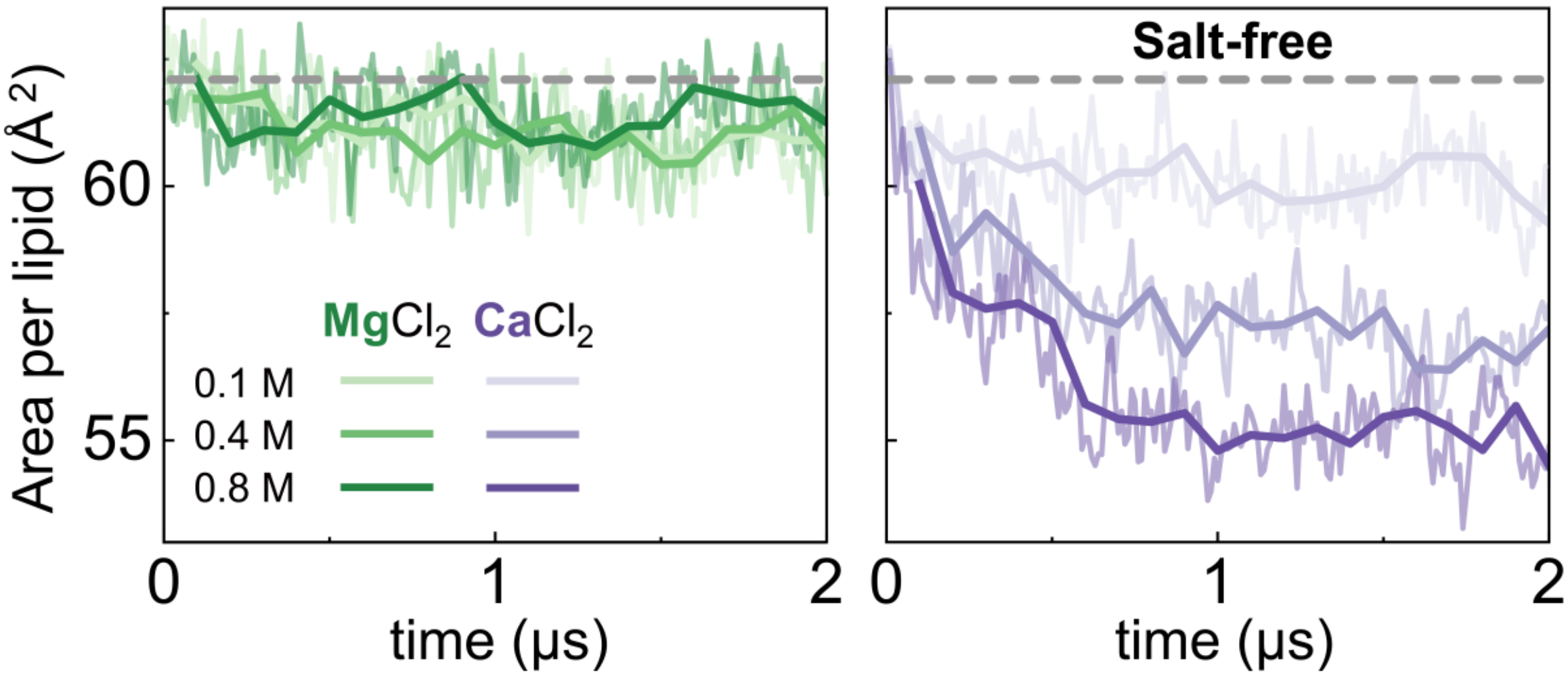

**FIG 2.** Time evolution of the area per lipid. (thin lines) Block averages of the raw data, recorded every 10 ps, over 10 ns period. (thick lines) Block averages of the raw data over 100 ns period.

After the equilibration procedure, production runs were carried out for our systems at 318 K under constant *NVT* conditions with the following procedures:

(1) To calculate the mean square displacement (MSD) and non-Gaussian parameter (NGP) for the lateral displacement of water molecules, shown in Figs. 3(a) and 3(c), trajectory data were recorded at two different sampling intervals. For the first 10 ns of the production run, trajectories were saved every 10 fs to investigate the short-time dynamics of water molecules. Subsequently, trajectories were saved every 1 ps for the following 1 μs to investigate the long-time dynamics of the systems.

(2) To obtain the layer-dependent profile of the lateral diffusion coefficient, $D_{\parallel}(n)$, shown in Fig. 4(c), we conducted additional simulation runs using umbrella sampling[42,57]. The simulation trajectories were recorded every 10 ps during the 1 μs-long *NVT* simulations. A more detailed description of this simulation process is presented in Appendix B.

Using the trajectories obtained from the production runs, we examine how salt species and their concentration influence the dynamics of water molecules confined between lipid bilayers. To quantify these dynamical effects, we first classified the water molecules according to their dynamical behavior. For this purpose, we first analyzed the time-averaged MSD of individual water molecules. We then calculated the ensemble-averaged MSD and NGP after excluding water molecules that exhibit long-time subdiffusive dynamics. The excluded water molecules are those trapped in the interfacial region through strong coordination with cations adsorbed on the membrane surface and exhibit dynamical behavior far distinct from that of most water molecules. Although the number of such trapped water molecules increases with salt concentration, they account for no more than 2.1 % of the total water population (see Table I). Details of the identification are provided in Appendix D.

# IV. Results and Discussion

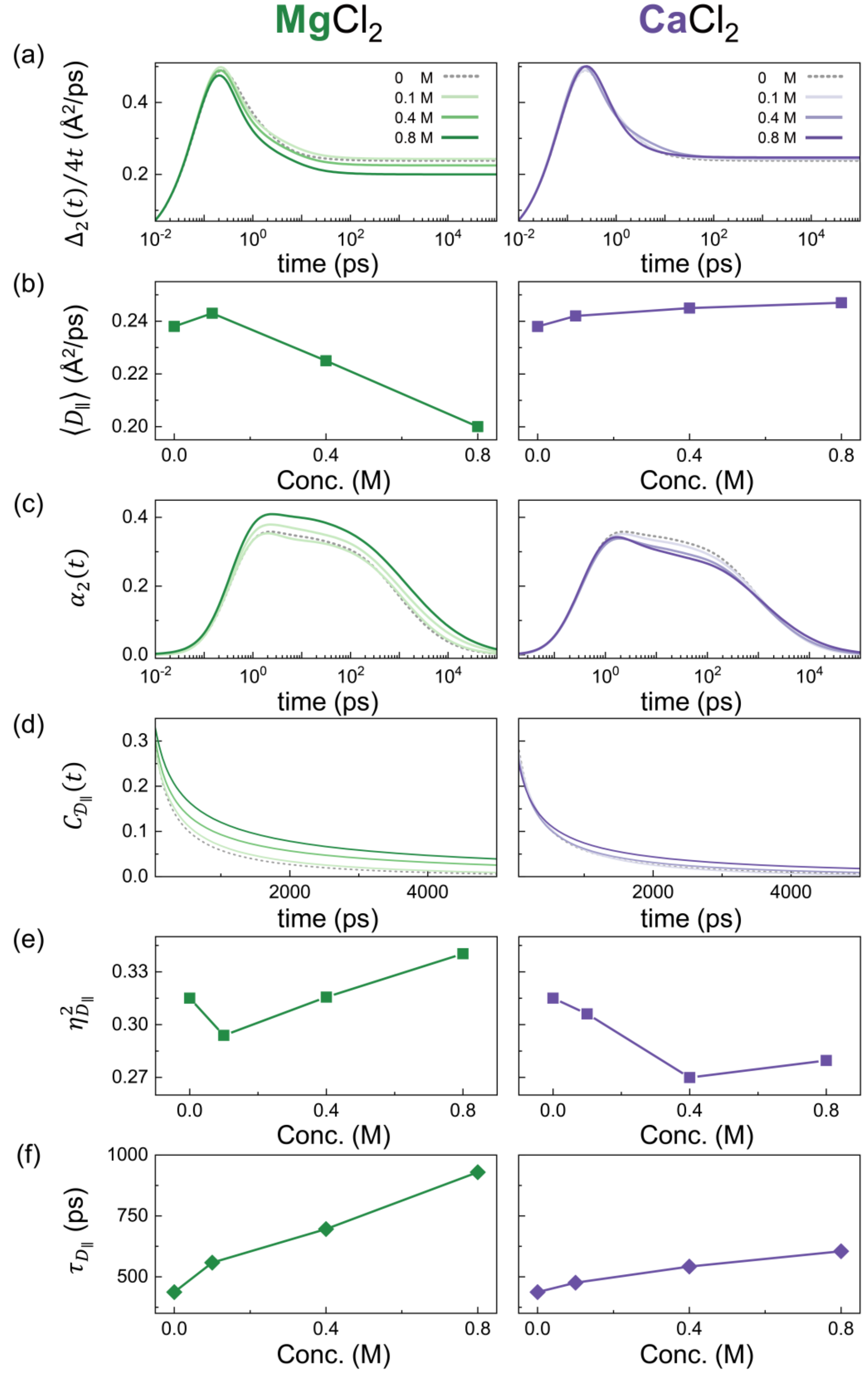

**FIG 3.** Transport dynamics of intermembrane water molecules in $MgCl_2$ and $CaCl_2$ solutions at various salt

concentrations. (a) Mean square displacement (MSD) divided by $\boldsymbol{4t}$, $\Delta_2(\boldsymbol{t})/\boldsymbol{4t}$, (b) the mean diffusion coefficient, $\langle \boldsymbol{D}_{\parallel} \rangle$ (c) Non-Gaussian parameter (NGP), $\boldsymbol{\alpha}_2(\boldsymbol{t})$, of lateral displacement distribution of water molecules. MSDs and NGPs are calculated from the MD simulation trajectories of water molecules. The mean diffusion coefficients are obtained from the long-time profile of the MSD. (d) Diffusion Kernel Correlations (DKC) extracted from the time profiles of the MSD and NGP using Eqs. (2) and (3). (e) Relative variance, $\boldsymbol{\eta}^2_{\boldsymbol{D}_{\parallel}}$, of the lateral diffusion coefficient and (f) the relaxation time, $\boldsymbol{\tau}_{\boldsymbol{D}_{\parallel}}$, of the diffusion coefficient fluctuation. $\boldsymbol{\eta}^2_{\boldsymbol{D}_{\parallel}}$ and $\boldsymbol{\tau}_{\boldsymbol{D}_{\parallel}}$ can be estimated from the maximum value of the DKC and the time integral of the DKC over the entire time domain, respectively (see the main text).

The diffusion coefficient exhibits a qualitatively different dependence on the concentrations of $Mg^{2+}$ and $Ca^{2+}$. In $CaCl_2$ solutions, $\langle D_{\parallel} \rangle$ monotonically increases with the salt concentration, whereas in $MgCl_2$ solutions, $\langle D_{\parallel} \rangle$ exhibits a non-monotonic dependence, largely decreasing with the salt concentration. The value of $\langle D_{\parallel} \rangle$ could be estimated by the long-time limit value of $\Delta_2(t)/4t$. For both salt solutions, the MSD, $\Delta_2(t)$, of water molecules reflects the dynamical transitions of water molecules' thermal motion [$\Delta_2(t) \sim t^2$] to terminal Fickian diffusion [$\Delta_2(t) \sim t^1$], through an intermediate sub-diffusive motion [$\Delta_2(t) \sim t^{\alpha}$ with $0 < \alpha < 1$] [see Fig. 3(a)], which is commonly observed across diverse complex fluids[41]. However, the diffusion coefficient and the overall time-profile of the MSD of water molecules exhibit a distinct dependence on the concentrations of $Mg^{2+}$ and $Ca^{2+}$. We note that only the short-time ballistic regime shows little dependence on salt concentration or cation species and exhibits quadratic time dependence, following $\Delta_2(t) = 2k_B T t^2/M$, where $M$, $k_B$, and $T$ denote the mass of a water molecule, the Boltzmann constant, and the temperature, respectively.

The NGP of the water displacement distribution also exhibits qualitatively distinct dependence on the concentrations of $Mg^{2+}$ and $Ca^{2+}$. The peak height of the NGP tends to increase with the salt concentration for $MgCl_2$ solutions but decreases with the salt concentration for $CaCl_2$ solutions [see Fig. 3(c)]. For both solutions, the relaxation time of the NGP increases with salt concentrations. Likewise, the long-time relaxation of the DKC, extracted from the MSD and NGP using Eqs. (2) and (3), occurs on a time scale that increases with salt concentrations for both solutions [see Fig. 3(f)]. From the time profile of the DKC, we extract the relative variance, $\eta^2_{D_\parallel}$, of the diffusion coefficient and the time correlation function, $\phi_{D_\parallel}(t)$, of the diffusion coefficient fluctuations. As the DKC reduces to $\eta^2_{D_\parallel}\phi_{D_\parallel}(t)$ at times longer than Fickian-diffusion onset time, the relative variance, $\eta^2_{D_\parallel}$, of the diffusion coefficient can be estimated by the peak height of the DKC time profile. The normalized time correlation function $\phi_{D_\parallel}(t)$ can then be extracted by scaling the DKC time profile by its peak value. The relaxation time, $\tau_{D_\parallel}$, of the diffusion coefficient fluctuation is defined by $\phi_{D_\parallel}(\tau_{D_\parallel}) = e^{-1}$. Our analyses clearly show that the value of $\tau_{D_\parallel}$ increases with salt concentrations for both salts. This result indicates that water molecules require a longer time to span the entire intermembrane space[42] as the salt concentrations increase. Notably, this slowing of the relaxation dynamics is more pronounced for $MgCl_2$ than for $CaCl_2$. In contrast, the relative variance $\eta^2_{D_\parallel}$ exhibits ion-specific behavior. The value generally increases as the concentration of $MgCl_2$ in solution rises, whereas it typically decreases with increasing $CaCl_2$ concentration. This is consistent with experimental observations[25].

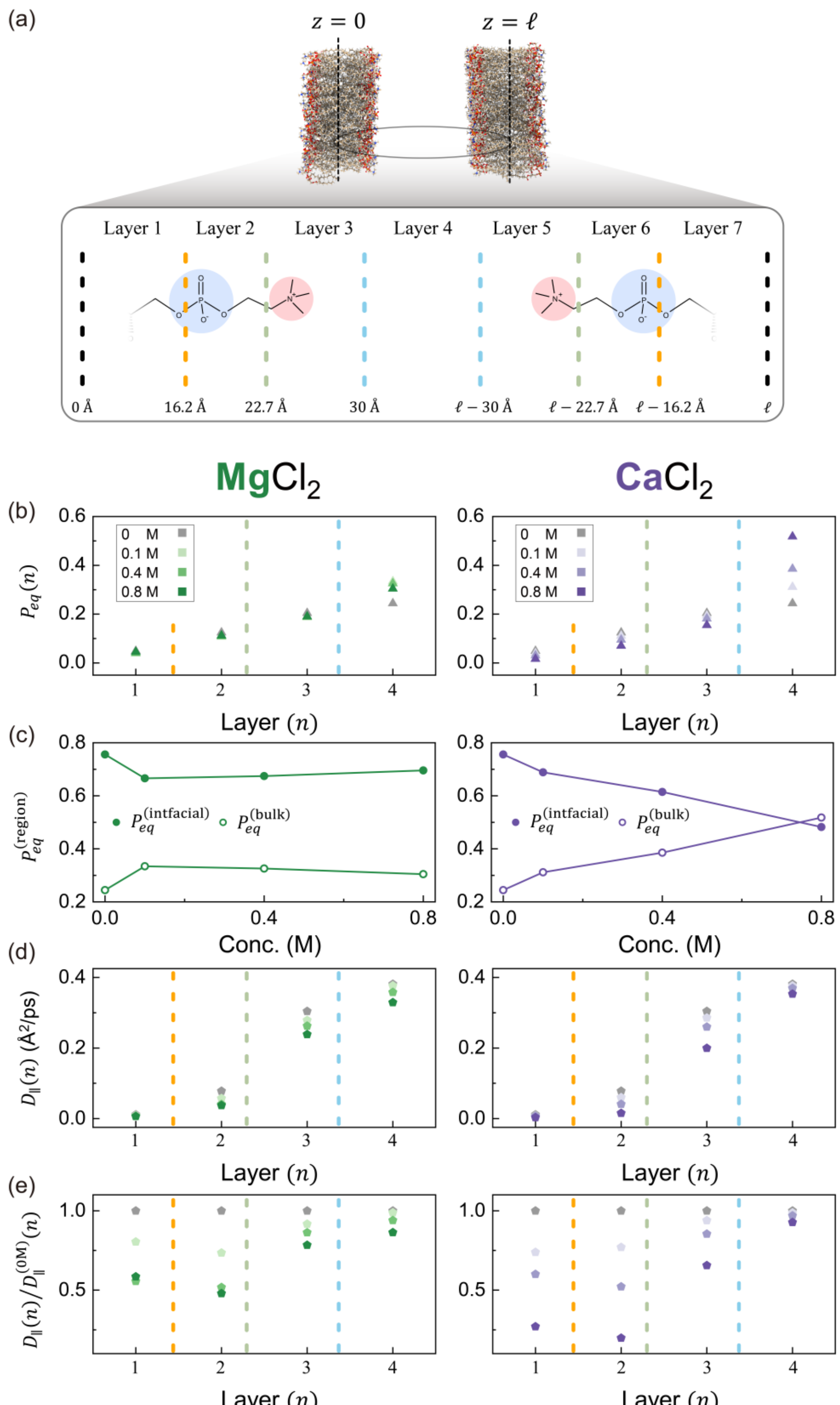

**FIG 4.** Effect of cation species and concentration on the water population distribution and lateral diffusion coefficient of water molecules at four different layers in the intermembrane space. (a) Schematic illustration of the layer definition. The vertical dashed lines, also shown in Fig. 7, indicate the boundaries between adjacent

layers. (b) Equilibrium population of water molecules at each layer at various salt concentrations. (c) (filled circles) Salt-concentration dependent population fraction $P_{eq}^{(\text{interfacial})}$ of water molecule at interfacial regions (layers 1-3 and 5-7) (empty circles) Salt-concentration dependent population fraction, $P_{eq}^{(\text{bulk})}\left(= 1 - P_{eq}^{(\text{interfacial})}\right)$, of water molecules at the bulk region (layer 4). (d) Lateral diffusion coefficient $D_{\parallel}(n)$ of water molecules at each layers for various salt concentrations. (e) Lateral diffusion coefficient $D_{\parallel}(n)$ of water molecules in ionic solution scaled by lateral diffusion coefficient $D_{\parallel}^{(0\text{M})}(n)$ of water molecules in the salt-free system. Layer 1-3 are equivalent to layers 7-5, due to the structural symmetry of our system.

The divalent cation-specific dependence of water transport dynamics on the salt concentrations can be understood in terms of the spatial variation of the lateral diffusion coefficient and water density along the $z$-axis. For this purpose, we discretize the intermembrane space into 7 distinct layers [see Fig. 4(a)] with different distances from the membrane's COM on the basis of the structure of water molecules shown in Figure 7, about which a more detailed discussion will be presented in Appendix A and B. Layer 4 at the center of our system, called bulk region, has the greatest water population and the water population symmetrically decreases with distance from the center. That is, the water population distribution across layer 1-3 on the left side of the system is the same as the water population distribution across layer 7-5 on the right side, which are called interfacial regions. The population fraction of water molecules and the water diffusion coefficient at each layer depend on both ion species and concentrations [Figs. 4(b) to 4(e)].

The population fraction of water molecules in the bulk region, represented by layer 4, is higher in the ionic solutions than in the salt-free system. This increase of bulk water population is particularly pronounced in the solution of $CaCl_2$ [Figs. 4(b) and 4(c)], which originates from the $Ca^{2+}$ ion induced increase in the lipid density that effectively squeezes the water molecules out of the membrane regions [Figs. 2 and 7(b)]. Due to the squeeze-out effect, the interfacial water population, $P_{eq}^{(\text{interfacial})}$ decreases with the $CaCl_2$ concentration [Fig. 4(c)]. In comparison, such squeeze-out effect is far smaller in the solution of $MgCl_2$ [see Figs. 2 and 7(b)] because $Mg^{2+}$ ions with larger hydration radii are less populated in lipid bilayer membranes than $Ca^{2+}$ ions with smaller hydration radii. Consequently, $Mg^{2+}$ is more abundant in the bulk region than $Ca^{2+}$ ions when the salt concentrations are the same [Fig. 7(g)]. The similar cation species dependent water population redistribution was experimentally reported in Fig. 3 of Ref. 25, where water molecules are categorized into tightly bound, loosely bound, and free water molecules. However, direct comparison between our simulation and the experimental results is not currently feasible for the lack of information about the spatial distribution of the three types of water molecules.

The distinct population distributions of $Ca^{2+}$ and $Mg^{2+}$ across interfacial and bulk regions cause water diffusion coefficients in each region to be dependent on ion species. Our MD simulation study shows that, in the absence of lipid membranes, the water diffusion coefficient in the $CaCl_2$ solution is similar to the water diffusion coefficient in the $MgCl_2$ solution if the two solutions have the same concentration (see Appendix C). However, for ionic biological water nanoconfined between lipid membranes, the $CaCl_2$ solution has a smaller water diffusion coefficient in the interfacial region than the $MgCl_2$ solution [see Figs. 4(d) and 4(e)]. This is because, in the interfacial region, $Ca^{2+}$ ions in the $CaCl_2$ solution are more populated than $Mg^{2+}$ ions in the $MgCl_2$ solution [Fig. 7(g)]. Consequently, the counter $Cl^-$ anions are also more abundant in the interfacial region in $CaCl_2$ solutions than in $MgCl_2$ solutions [Fig. 7(h)]. On the other hand, the diffusion coefficient of water molecules in the bulk region is smaller in $MgCl_2$ solutions than in $CaCl_2$ solutions [Figs. 4(d) and 4(e)] because $MgCl_2$ solutions have higher ion concentrations in the bulk region [Figs. 7(g) and 7(h)]. This ion-specific trend of water diffusion coefficient is also consistent with the QENS experiment in Ref. 25, where Seto and Yamada reported that the diffusion coefficient of free water is higher in $CaCl_2$ than in $MgCl_2$, whereas the diffusion coefficient of bound water is lower in $CaCl_2$ than in $MgCl_2$.

The salt concentration-dependent redistribution of ions and water populations, together with the resulting changes in the water diffusion coefficients in each region, provides an explanation for the salt concentration-dependent mean diffusion coefficient of water molecules. In $CaCl_2$ solutions, the water population in the bulk region, with the highest water diffusion coefficient, is far higher than in the interfacial regions due to the squeeze-out effects of $Ca^{2+}$ ion. Such water population shift to the bulk region increases almost linearly with salt concentration for $CaCl_2$ solutions [Fig. 4(c)], which explains the increase in the average water diffusion coefficient with the salt concentration for $CaCl_2$ solutions [Fig. 3(b)]. On the other hand, in $MgCl_2$ solutions, such increase in the mean diffusion coefficient due to the water population shift to the bulk region is observed only in 0.1 M solution [Fig. 3(b)]. When $MgCl_2$ concentration exceeds 0.1 M, the mean diffusion coefficient of water molecules decreases with salt concentration. This results because, when the $MgCl_2$ concentration is higher than 0.1 M, the water population distribution across different layers exhibits only a marginal dependence on $MgCl_2$ concentration, whereas the diffusion coefficient in each layer significantly decreases with salt concentration as water molecules interacting with ions have lower motility than free water molecules[58].

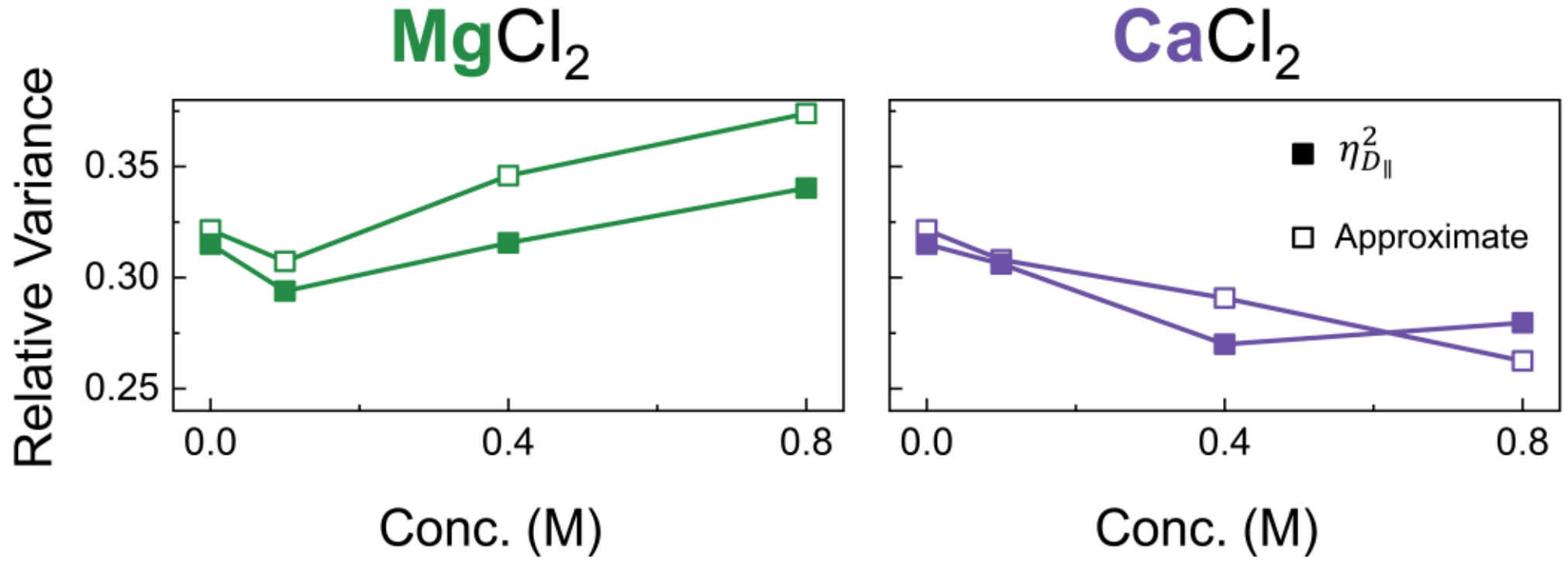

**FIG 5.** Salt-concentration dependence of the relative variance of the water diffusion coefficient in intermembrane $MgCl_2$ and $CaCl_2$ solutions. (filled squares) Results obtained from the peak height of the DKC. (empty squares) Approximate results using the layer-dependent values of water diffusion coefficients and water population fractions shown in Fig. 4.

The variation of the water diffusion coefficient and population along the longitudinal direction, or along the *z*-axis, serves as the major contributor to the relative variance of the water diffusion coefficient. The approximate estimation of the relative variance of the water diffusion coefficient, obtained from the layer-dependent water diffusion coefficient and population values, exhibits the similar magnitude and salt concentration dependence to the accurate relative variance values of the water diffusion coefficient, obtained from the peak height of the DKC (see Fig. 5).

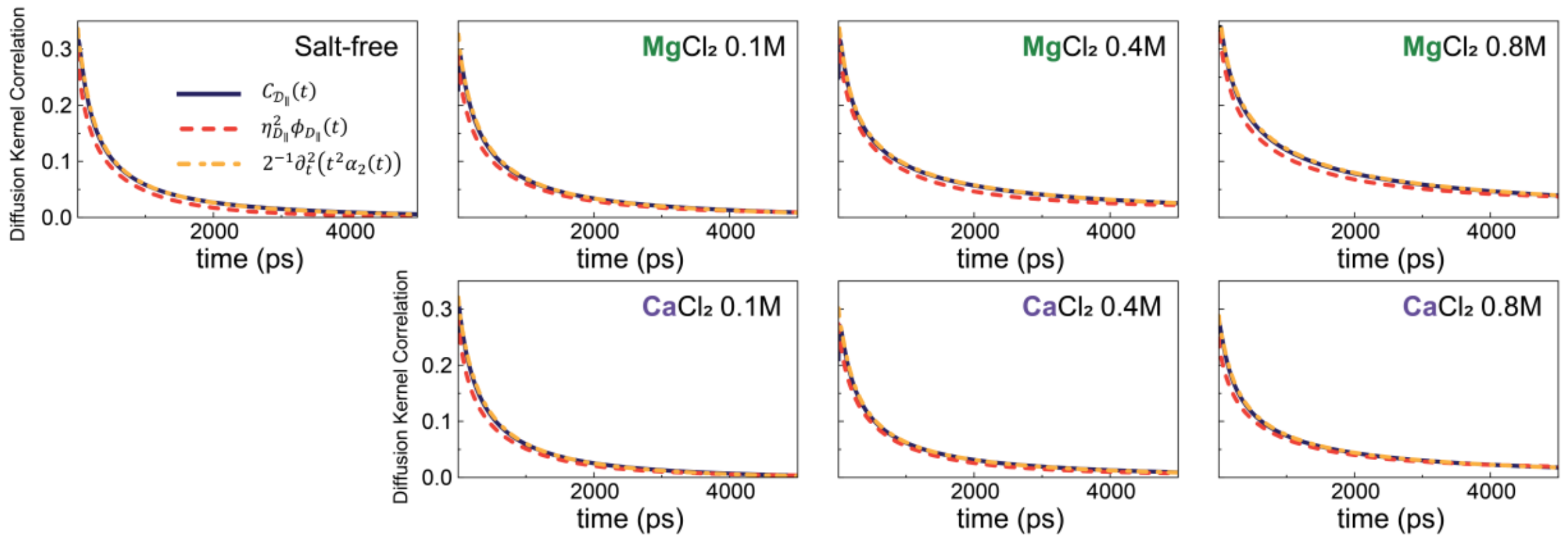

**FIG 6.** Time profiles of the diffusion kernel correlation (DKC) of water molecules at various concentrations of $MgCl_2$ and $CaCl_2$ solution confined between two lipid membranes. (solid lines) DKC time profile extracted from the MSD and NGP time profiles using Eqs. (2) and (3). (red dashed lines) Approximate DKC time profiles calculated from Eqs. (8) to (10). (yellow dash-dotted lines) the time correlation function of the diffusion coefficient fluctuations calculated by Eq. (7) and the NGP time profile.

The relaxation dynamics of the diffusion coefficient fluctuation can also be explained in terms of the variation of the water diffusion coefficient and population dynamics along the longitudinal direction. We find that both the long-time DKC extracted from the MSD and NGP time profiles and $\eta^2_{D_\parallel}\phi_{D_\parallel}(t)$ calculated from Eq. (7) are in good agreement with the following approximate expression (Fig. 6):

$$\eta^2_{D_\parallel}\phi_{D_\parallel}(t) \cong \langle D_\parallel \rangle^{-2} \sum_n \sum_m \delta D_\parallel(n) G^{(d)}(n,t|m) \delta D_\parallel(m) P_{eq}(m) \qquad (10)$$

where $G^{(d)}(n,t|m)$ denotes the conditional probability that a water molecule is found in layer $n$ at time $t$, given that the water molecule is initially located in layer $m$. The time profiles of $G^{(d)}(n,t|m)$ required for calculation of Eq. (10) were obtained from the trajectories of water molecules in the MD simulations. The agreement between the accurate time profile of $\eta^2_{D_\parallel}\phi_{D_\parallel}(t)$ and the approximate result, Eq. (10), shows that the relaxation dynamics of the diffusion coefficient fluctuation is governed by the variation of the water diffusion coefficient and transport dynamics of water molecules along the *z*-axis. The relaxation time of the diffusion coefficient fluctuation monotonically increases with the salt concentration for both ions [see Fig. 3(f)].

# V. Conclusions

We investigated the effects of divalent ions on the transport dynamics of water molecules nanoconfined between DMPC lipid bilayers in solutions of $CaCl_2$ and $MgCl_2$ by using all-atom MD simulations and the generalized transport equation of biological water. Our results show that the lateral diffusion of intermembrane water molecules exhibits a transient super-Gaussian displacement distribution whose characteristics depend on the divalent cation species and concentration. The mean lateral diffusion coefficient of biological water monotonically increases with $Ca^{2+}$ concentration, whereas it exhibits a largely opposite, non-monotonic dependence on $Mg^{2+}$ concentration. The transient deviation from a Gaussian also displays ion-specific behavior. By analyzing the MD simulation results for the mean square displacement and non-Gaussian parameter of the lateral displacement distribution using the solution of the generalized transport equation for biological water, we extracted the relative variance of the diffusion coefficient and the time-correlation function of the diffusion coefficient fluctuations. In $CaCl_2$ solution, the $Ca^{2+}$-induced decrease in the area per lipid effectively squeezes water molecules out of the interfacial region. As a result, the fraction of water molecules in the bulk region increases with $CaCl_2$ concentrations, where the diffusion coefficient is larger than in the interfacial regions. This leads to an increase in the mean diffusion coefficient and a decrease in the relative variance of the lateral diffusion coefficient with salt concentrations. In contrast, in $MgCl_2$ solution, $Mg^{2+}$ ions with larger hydration radii than $Ca^{2+}$ ions, penetrate into the lipid membrane far less than $Ca^{2+}$ ions do and induce far smaller population increase of water molecules in the bulk region than $Ca^{2+}$ ions. Instead, in $MgCl_2$ solution, ions are more concentrated in the bulk region than in $CaCl_2$ solution. Consequently, the lateral diffusion coefficient of water molecules in the bulk region is significantly smaller in $MgCl_2$ solution than that in $CaCl_2$ solution. This leads to an overall decrease in the mean diffusion coefficient with $MgCl_2$ concentration. On the other hand, the heterogeneity in the diffusion coefficient increases with salt concentrations in $MgCl_2$ solution. These results could be quantitatively explained by assuming that the primary source of the lateral diffusion coefficient fluctuation is thermal motion of water molecules in the longitudinal direction, along which microscopic environments surrounding a water molecule, including the functional groups of lipid membrane, ion concentrations, and hydrogen bond network drastically change.

# Acknowledgments

This work was supported by the Global Science Research Center Program funded by the National Research Foundation of Korea (RS-2024-00411134) and by the Chung-Ang University Graduate Research Scholarship in 2020. MC gratefully acknowledges financial support from the Institute for Basic Science, Korea (IBS-R023-D1). The authors declare no competing financial interest.

# Data Availability

The data that support the findings of this study are available from the corresponding author upon reasonable request.

# Appendix

Appendices include structural properties of the interfacial region along the $z$-axis (Appendix A), a detailed method for calculating layer-dependent diffusion coefficients (Appendix B), the salt concentration dependence of the water diffusion coefficient in bulk phase (Appendix C), the classification of water molecules into trapped and non-trapped groups (Appendix D), the rotational dynamics of interfacial water molecules (Appendix E), and the salt-induced water squeeze-out effect for cholesterol-incorporated DMPC bilayer membrane system (Appendix F).

## Appendix A: STRUCTURAL PROPERTIES OF INTERFACIAL REGION

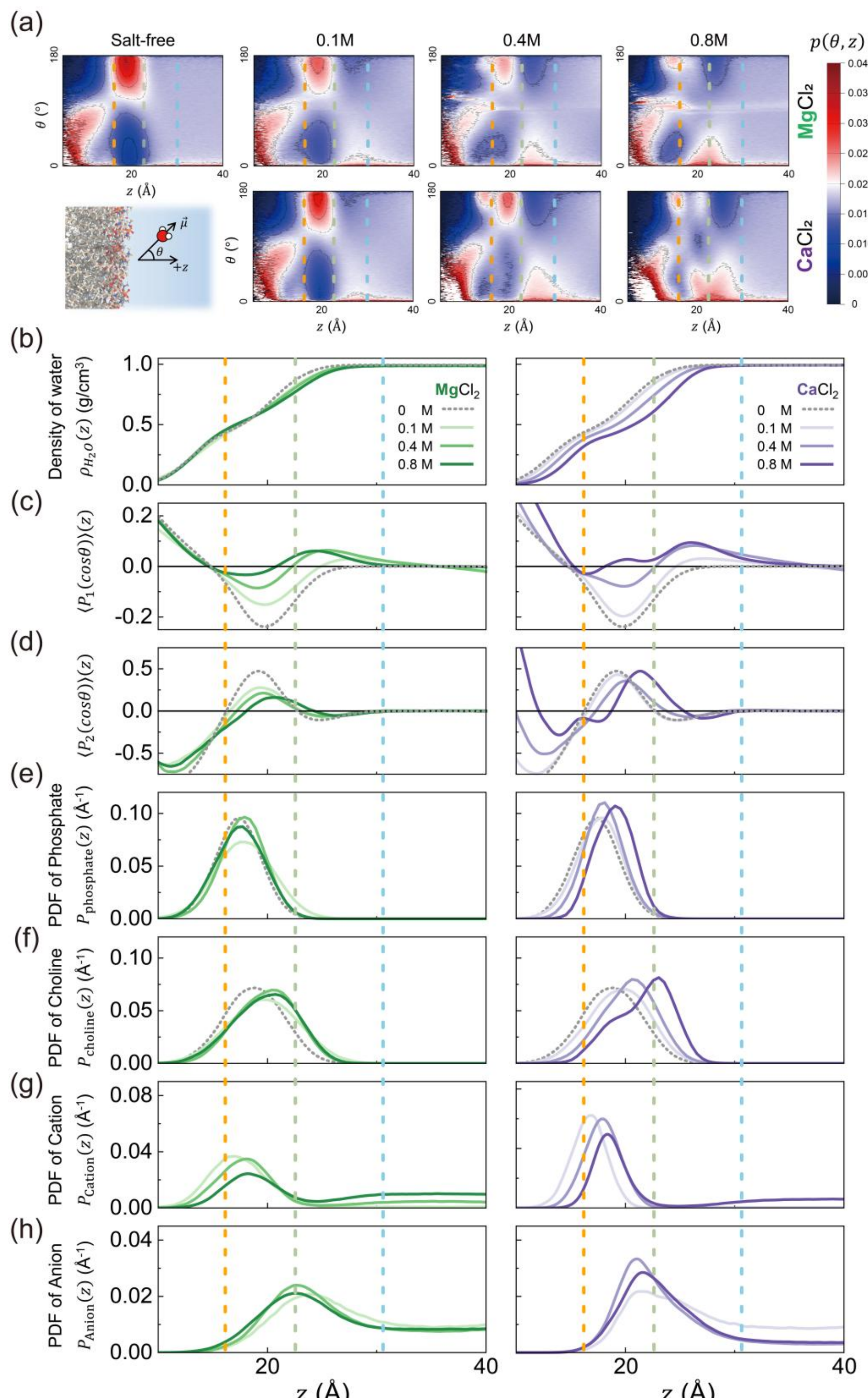

**FIG 7.** (a)-(d) Dependence of structural properties of the water molecules on the distance, *z*, from the COM of the lipid membrane: *z*-dependent profile of (a) probability distribution of water dipole alignments, $\boldsymbol{p(\theta, z)}$. $\boldsymbol{\theta}$ denotes the angle between the water dipole and the *z*-axis shown in the leftmost panel of the second row. (b) *z*-

dependent mass density profile, $\boldsymbol{\rho}_{\mathbf{H_2O}}(\boldsymbol{z})$, (c) the first-order Legendre polynomial, $\langle \boldsymbol{P_1}(\mathbf{cos}\,\boldsymbol{\theta})\rangle(\boldsymbol{z})$, representing the net dipole alignment along the surface normal, and (d) the second-order Legendre polynomial, $\langle \boldsymbol{P_2}(\mathbf{cos}\,\boldsymbol{\theta})\rangle(\boldsymbol{z})$, characterizing the degree of orientational ordering. (e-h) Probability distribution of functional groups in the DMPC and various ions along the *z*-axis (e-f) Probability distributions of phosphate and choline groups of the DMPC (g)-(h) Probability distributions of cations and anions. Using the zeros of $\langle \boldsymbol{P_2}(\mathbf{cos}\,\boldsymbol{\theta})\rangle(\boldsymbol{z})$ of the salt-free (0M) system, we divided the intermembrane space into seven discrete layers. The layer boundaries are indicated by dotted lines in (b)-(h), located at *z* = 16.2, 22.7, and 30 Å.

To investigate the effects of salt ions on the structure of water molecules near the lipid bilayer, we calculate several structural properties along the *z*-axis. Figure 7(a) shows the *z*-dependence of the angular distribution of water dipole alignments[15,59,60], $p(\theta,z)$, for various salt concentrations and cation species, which satisfies the following normalization condition: $\int_0^\pi d\theta \sin\theta\, p(\theta,z) = 1$. Here, $\theta$ and *z* respectively denote the angle between the water dipole moment and the surface vector of the lipid bilayer, or the *z*-axis, as illustrated in the lower-left panel of Fig. 7(a), and the distance between the COM of water molecule and the COM of the bilayer, as illustrated in Fig. 1. From $p(\theta,z)$, we compute the average of the first- and second-order Legendre polynomials, $\langle P_1(\cos\theta)\rangle(z) = \langle\cos\theta\rangle(z)$ and $\langle P_2(\cos\theta)\rangle(z) = \langle(3\cos^2\theta - 1)/2\rangle(z)$ which characterize the extent and direction of dipole alignment along the *z*-axis, as shown in Figs. 7(c) and 7(d). Here, bracket $\langle \ldots \rangle(z)$ denotes the average over the angular distribution of water molecules located within an interval ($z - 0.125$ Å, $z + 0.125$ Å), i.e., $\langle P_n(\cos\theta)\rangle(z) = \int_0^\pi d\theta \sin\theta\, P_n(\cos\theta) p(\theta,z)$ . $\langle P_1(\cos\theta)\rangle(z)$ provides information about whether the average water dipole moment points toward the +*z* or -*z* direction, whereas $\langle P_2(\cos\theta)\rangle(z)$ provides information about whether orientation of water dipole tends to be in parallel or perpendicular to the lipid bilayer surface[39].

For the salt-free case, water molecules with dipoles oriented toward the membrane surface ($\theta$ = 180°) are dominantly distributed between 16.2 Å and 22.7 Å from the membrane's COM, which corresponds to the first and second zero-crossing points of $\langle P_2(\cos\theta)\rangle(z)$ in the salt-free case or the orange dotted and light-green dotted lines in Fig. 7. This preferential orientation arises because phosphate groups [Fig. 7 (e)], strongly constrain the orientation of nearby water molecules through hydrogen bonding, whereas choline groups [Fig. 7 (f)] do not have such strong hydrogen bonding with water molecules because the methyl groups in the choline groups sterically hinder water molecules from interacting directly with the positively charged nitrogen[42,61–64]. For $z$ < 16.2 Å, the dipoles of water molecules tend to align perpendicular to the membrane surface vector, ($\theta$ = 90°), which reflects the dominant orientational influence of the carbonyl groups in this region, whose dipoles also align perpendicular to the membrane surface [15].

As salt concentration increases, more ions are adsorbed onto the membrane surface, as shown in Figs. 7(g) and 7(h). This results in ionic screening of interactions between water molecules and charged headgroups in the lipid membranes and weakens the orientational ordering of water molecules located near phosphate and choline groups, which are located within the orange and light-green dotted lines in Fig. 7. This decrease in the orientational ordering of water molecules is represented by the gradual fading of red and blue colors in Fig. 7(a). This trend is confirmed in the ion-concentration dependence of the first- and second-order Legendre polynomials shown in Figs. 7(c) and 7(d). Note that the degree of this weakening depends strongly on the identity of the cation. For $MgCl_2$, increasing salt concentration leads to a systematic decrease in the magnitudes of $\langle P_1(\cos\theta)\rangle(z)$ and $\langle P_2(\cos\theta)\rangle(z)$, indicating that the average dipole alignment loses its preferred orientation. In contrast, for $CaCl_2$, a sign reversal of $\langle P_1(\cos\theta)\rangle(z)$ is observed, while the sign and magnitude of $\langle P_2(\cos\theta)\rangle(z)$ remain preserved with increasing salt concentration. This behavior indicates an inversion of the average dipole orientation along the $z$-axis. This result is consistent with experimentally observed trends that are obtained from vibrational sum frequency generation (VSFG) spectra detecting OH stretching vibration and direction of dipole moment of interfacial water[22,65]. It is notable that these results might appear that the interaction between membrane surface and water molecules decreases as the ion concentration increases. However, the diffusion coefficient in the interfacial region [see Figs. 4(d) and 4(e)] shows that the diffusive motion of water molecules in this region is slowed down. This observation suggests that the direct membrane-water interaction is replaced by an ion-mediated interaction between the membrane and water.

This contrast reflects the distinct roles of the two cations at the membrane surface. $Ca^{2+}$ efficiently screens the effective headgroup charge, dominated by the phosphate groups, thereby inducing a reversal of the average dipole orientation. In contrast, $Mg^{2+}$ remains strongly coordinated with its hydration shell, which limits its ability to effectively screen the membrane surface charge[15,66]. This difference in interfacial binding is also reflected in the area per lipid (APL) shown in Fig. 2. While $MgCl_2$ does not lead to a noticeable reduction in APL, $CaCl_2$ induces a gradual decrease in APL with increasing concentration, consistent with enhanced lipid packing driven by direct $Ca^{2+}$–headgroup coordination. This cation-specific difference in lipid packing is further manifested in the water density profile, $\rho_{\mathrm{H_2O}}(z)$, shown in Fig. 7(b). While the interfacial water density profile shows little variation for $MgCl_2$, $CaCl_2$ leads to a pronounced reduction of $\rho_{\mathrm{H_2O}}(z)$ in the interfacial region as the concentration increases, consistent with the squeezing out of interfacial water due to bilayer condensation.

Beyond 22.7 Å, $\langle P_1(\cos\theta)\rangle(z)$ remains negative and approaches 0 monotonically in the salt-free case. With adding salt, however, $\langle P_1(\cos\theta)\rangle(z)$ in this region becomes positive, and the relaxation toward 0 becomes more gradual. The positive $\langle P_1(\cos\theta)\rangle(z)$ in this region indicates that cations adsorbed onto the membrane surface influence the orientation of water molecules via long-range electrostatic interactions. Considering the corresponding radial distribution functions between oxygen atom of water and cation (See Fig. 8) and the distribution of cation [see Fig. 7(g)], this contribution originates from water molecules located beyond the first hydration layer. However, this interaction is not sufficiently strong to induce rigid orientational ordering, as shown in $\langle P_2(\cos\theta)\rangle(z)$ in this region. Note here that, irrespective of salt species and concentration, $\langle P_2(\cos\theta)\rangle(z)$ converges to zero beyond approximately 30 Å (light blue dotted line in Fig. 7). Beyond 30 Å, the density profile also recovers its respective bulk-limit value.

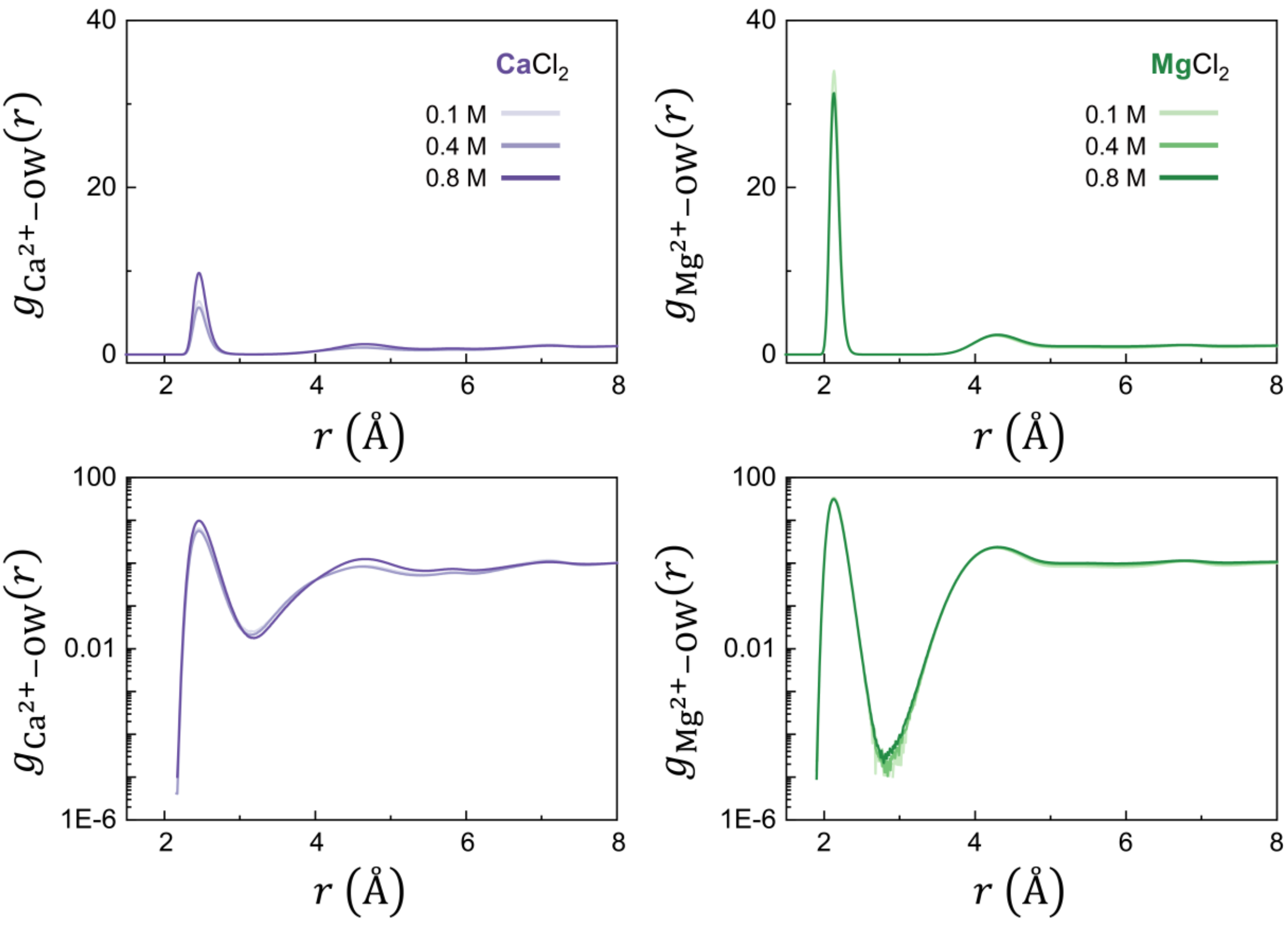

FIG **8.** Radial distribution functions (RDFs) of divalent cations and the oxygen atoms of water molecules.

From the zero-crossing points of $\langle P_2(\cos\theta)\rangle(z)$ obtained from the salt-free (0M) system and the region where $\langle P_2(\cos\theta)\rangle(z)$ approaches zero [see Fig. 7(d)], we divided the intermembrane space into seven discrete layers. Based on this layer definition, we analyzed the population and the lateral diffusion coefficient of water molecules in each layer, as shown in Fig. 4. The population, $P_{eq}(n)$, of each layer is obtained by integrating the mass density profile, $\rho_{\mathrm{H_2O}}(z)$, of water molecules [Fig. 7(b)] over each layer and normalizing by the total population. Using umbrella sampling (see section III and Appendix B), we evaluated the corresponding layer-dependent diffusion coefficients, $D_{\parallel}(n)$ [see Fig. 4(d)]. Among the seven layers, layer 4 represents the bulk region, whereas layers 1 to 3 are equivalent to layers 7 to 5, respectively.

## Appendix B. METHOD FOR CALCULATING LAYER-DEPENDENT DIFFUSION COEFFICIENTS

To determine the layer-dependent profile of the lateral diffusion coefficient, we first evaluate the layer-resolved MSD of water molecules in the intermembrane space. For this purpose, we compute the MSD time profile of water molecules initially located within each layer. It is difficult to estimate the long-time behavior of the MSD accurately, because very few water molecules remain in the initial layer at long times[57]. To overcome this difficulty, we perform a constrained MD simulation for the first 4 layers. In these simulations, approximately 10 % of the water molecules within a given layer are randomly selected and subjected to an external potential, $U_n(z)$, defined as

$$U_n(z) = \begin{cases} 0 \left(z_{\min}^{(n)} < z < z_{\max}^{(n)}\right) \\ k\left(z - z_{\min}^{(n)}\right)^2 \left(z < z_{\min}^{(n)}\right) \\ k\left(z - z_{\max}^{(n)}\right)^2 \left(z > z_{\max}^{(n)}\right) \end{cases} . \tag{11}$$

Here, $U_n(z)$ denotes the restraining potential applied to water molecules in the *n*-th layer, and $z_{\min}^{(n)}$ and $z_{\max}^{(n)}$ represent the lower and upper boundaries of that layer, respectively. This restraining potential prevents the selected molecules from crossing the layer boundaries. Here, $k$ represents the spring constant whose value is set to be 1.25 [Kcal/(mol·Å$^2$)]. In the constrained MD simulation, the MD trajectories were recorded every 10 ps during 1 μs-long *NVT* simulations. From the trajectories of water molecules constrained by the harmonic potential, we obtained MSDs of the lateral water displacement for every layer of the simulation system. From the long-time MSD profile, we estimate the lateral diffusion coefficient at each layer by $D_{\parallel}(n) = \lim_{t\to\infty} \Delta_2(n,t)/4t$. These values are shown in Fig. 4(d).

Using $D_{\parallel}(n)$, we approximately calculated $\langle D_{\parallel}^{m}\rangle$ by

$$\left\langle D_{\parallel}^{m}\right\rangle = \int_0^{\ell} dz\ D_{\parallel}^{m}(z) p_{eq}(z) \cong \sum_{n=1}^{7} D_{\parallel}^{m}(n) P_{eq}(n) \tag{12}$$

with $p_{eq}(z) = \rho_{\mathrm{H_2O}}(z) / \int_0^{\ell} dz' \rho_{\mathrm{H_2O}}(z')$ and $P_{eq}(n) = \int_{z_{\min}^{(n)}}^{z_{\max}^{(n)}} dz\, p_{eq}(z)$. Here, $\rho_{\mathrm{H_2O}}(z)$ denotes the mass density profile of water molecules, shown in Fig 7 (b). Using this $\langle D_{\parallel}^{m} \rangle$, the relative variance of the water diffusion coefficient can be approximately estimated from Eq. (9), as shown in Fig. 5.

## Appendix C. SALT CONCENTRATION DEPENDENCE OF THE WATER DIFFUSION COEFFICIENT IN BULK PHASE

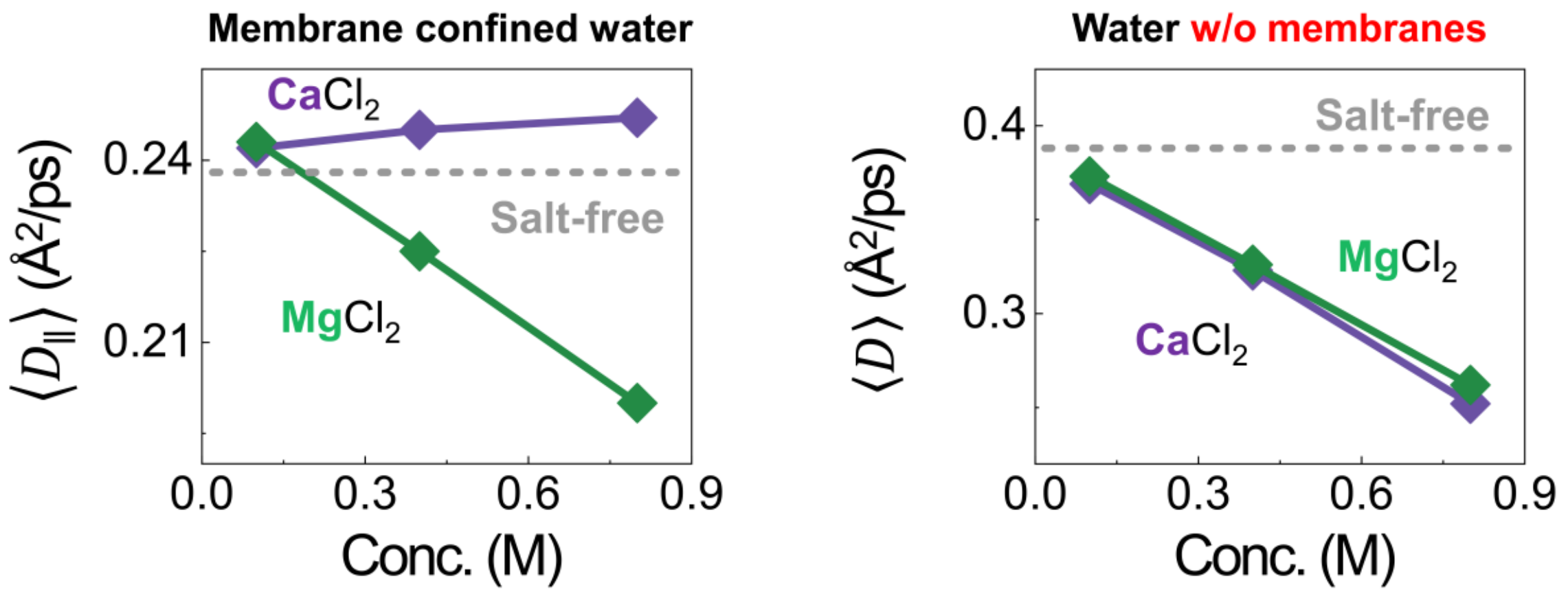

**FIG 9.** The lateral diffusion coefficients of water molecules in the membrane-confined ionic solutions and the diffusion coefficient of water molecules in the membrane-free solutions at various salt concentrations.

To obtain the diffusion coefficient of water molecules in bulk solution, we performed an additional set of simulations. The simulation systems were constructed using CHARMM-GUI solution builder[48,50,67], where the initial cubic box size was set to $60 \times 60 \times 60$ Å$^3$. The resulting simulation box contained approximately 6,000 water molecules.

The MD simulation was carried out using the same force field and simulation parameters as those used in the main simulations. The initial configuration was first stabilized through energy minimization for 5,000 steps, using a combination of the steepest descent and the conjugate gradient methods. This was followed by a 2 ns *NpT* equilibration at 1.0 atm and 318 K, using isotropic position scaling with a relaxation time of 2 ps. Temperature control was achieved using a Langevin thermostat with a collision frequency of 1.0ps$^{-1}$. Subsequently, a 2 ns *NVT* simulation at 318 K was performed using the Langevin thermostat to ensure that the system reached thermal equilibrium. After the equilibration procedure, production runs were carried out at 318 K under constant *NVT* conditions. Trajectories were saved every 10 ps during 1 μs-long *NVT* simulations.

From the trajectories of water molecules, the three-dimensional mean square displacements were calculated, and the diffusion coefficients, $\langle D \rangle$, of water molecules in bulk solutions were obtained from their long-time linear regimes. The diffusion coefficients, $\langle D \rangle$, of water molecules in various conditions are shown in Fig. 9. For comparison, the lateral diffusion coefficients, $\langle D_{\parallel} \rangle$, of membrane-confined water molecules discussed in the main text [see Fig. 3(b)] are also included in this figure.

## Appendix D. CLASSIFICATION OF WATER MOLECULES INTO TRAPPED AND NON-TRAPPED GROUPS

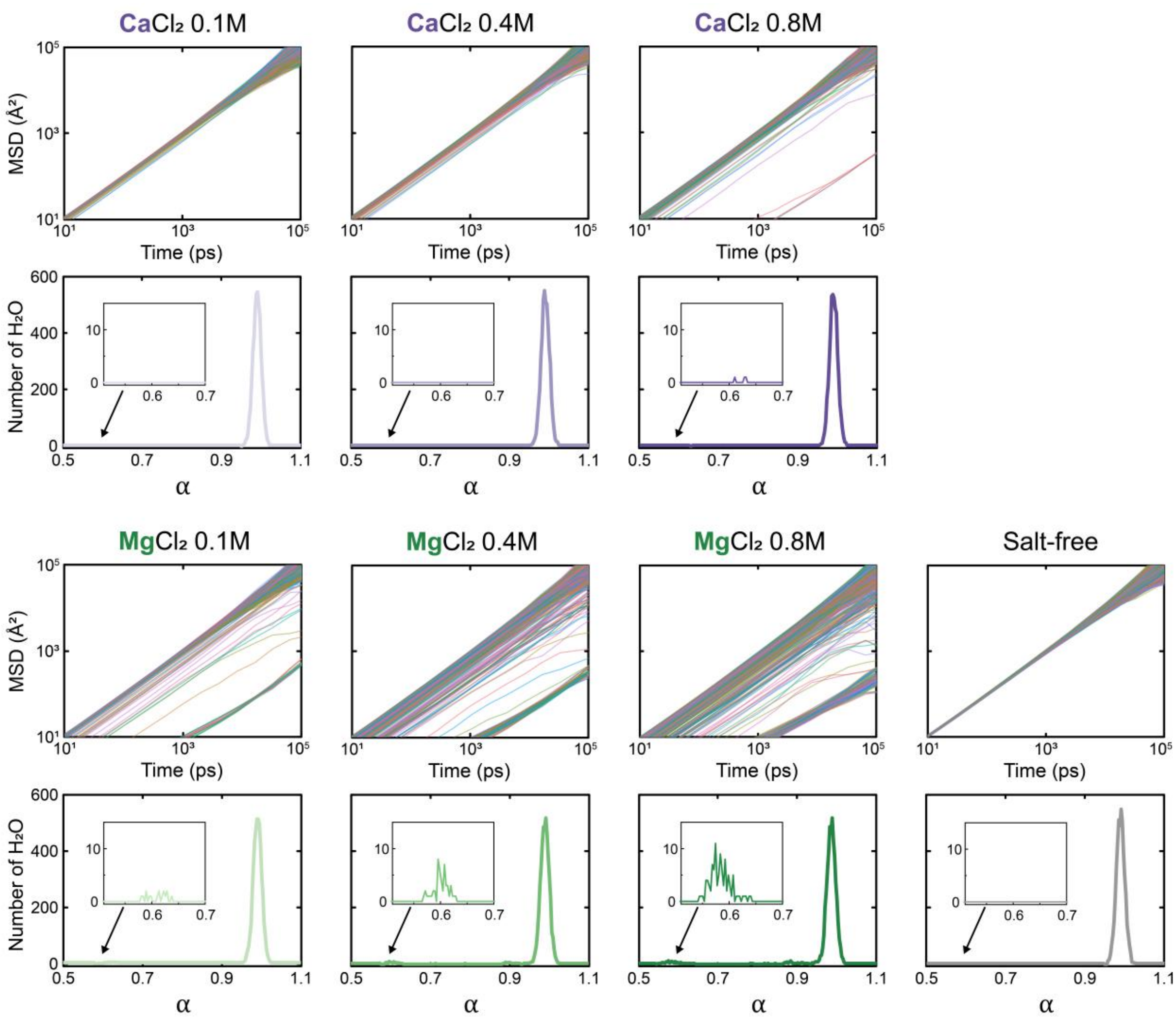

**FIG 10.** Time-averaged MSD of each individual water molecule at various salt concentrations in $CaCl_2$ solution (first row) and in $MgCl_2$ solution (third row). The distribution of exponent $\boldsymbol{\alpha}$ defined by $\boldsymbol{\alpha \equiv \partial \ln \Delta_2(t)/\partial \ln t}$ at 100 ps for $CaCl_2$ solutions (second row) and for $MgCl_2$ solutions (fourth row).

As shown in Fig. 10, we obtained time-averaged MSD of individual water molecules. From this result, we excluded water molecules that exhibit long-time subdiffusive behavior. More specifically, we examined the MSD at $t$ = 100 ps and excluded molecules whose MSD follows $\sim t^{\alpha}$ with an exponent $\alpha$ lower than 0.75 (see Table I).

| | Number of water molecules | | Percentage of trapped water molecules (%) |
|---|---|---|---|
| | non-trapped | trapped | |
| Salt-free | 4736 | 0 | 0 |
| $CaCl_2$ 0.1M | 4736 | 0 | 0 |
| $CaCl_2$ 0.4M | 4736 | 0 | 0 |
| $CaCl_2$ 0.8M | 4732 | 4 | 0.084 |
| $MgCl_2$ 0.1M | 4720 | 16 | 0.34 |
| $MgCl_2$ 0.4M | 4684 | 52 | 1.10 |
| $MgCl_2$ 0.8M | 4638 | 98 | 2.07 |

Table I. Total number of water molecules, number of trapped water molecules, and percentage of trapped water molecules.

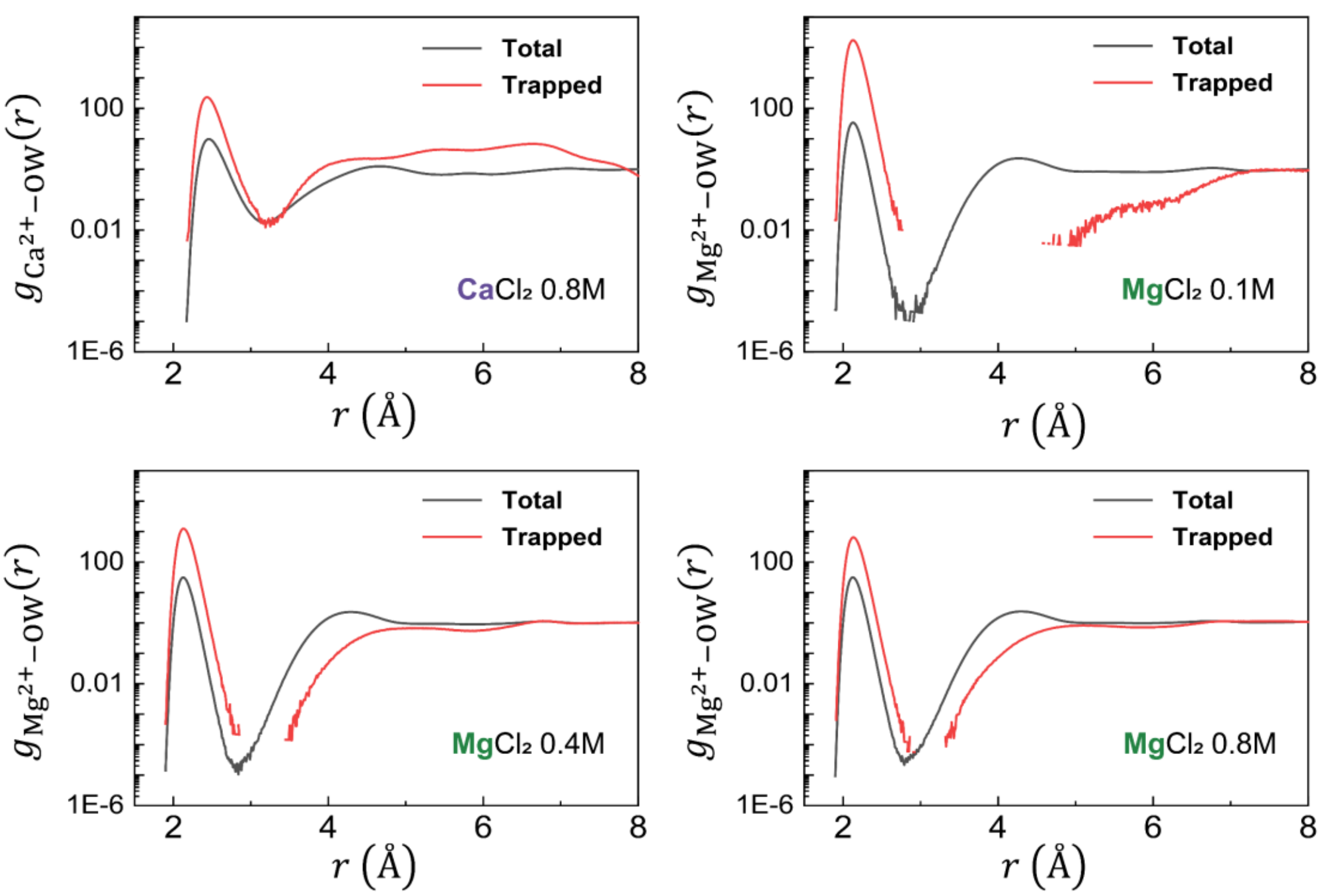

**FIG 11.** Radial distribution functions (RDFs) of cations and the oxygen atoms of trapped water molecules (red lines). The RDFs of cations and total water molecules (black lines)

To investigate the origin of this persistent subdiffusion, we computed the radial distribution function (RDF) between these water molecules and salt cation, as shown in Fig. 11. The results reveal that water molecules exhibiting long-time subdiffusion interact strongly with cations and rarely escape from the cation hydration shell. Compared with the RDF obtained from the remaining water molecules, the RDF of the subdiffusive molecules shows a significantly enhanced first-shell peak (see Fig. 11). This indicates that these water molecules are strongly trapped by nearby cations.

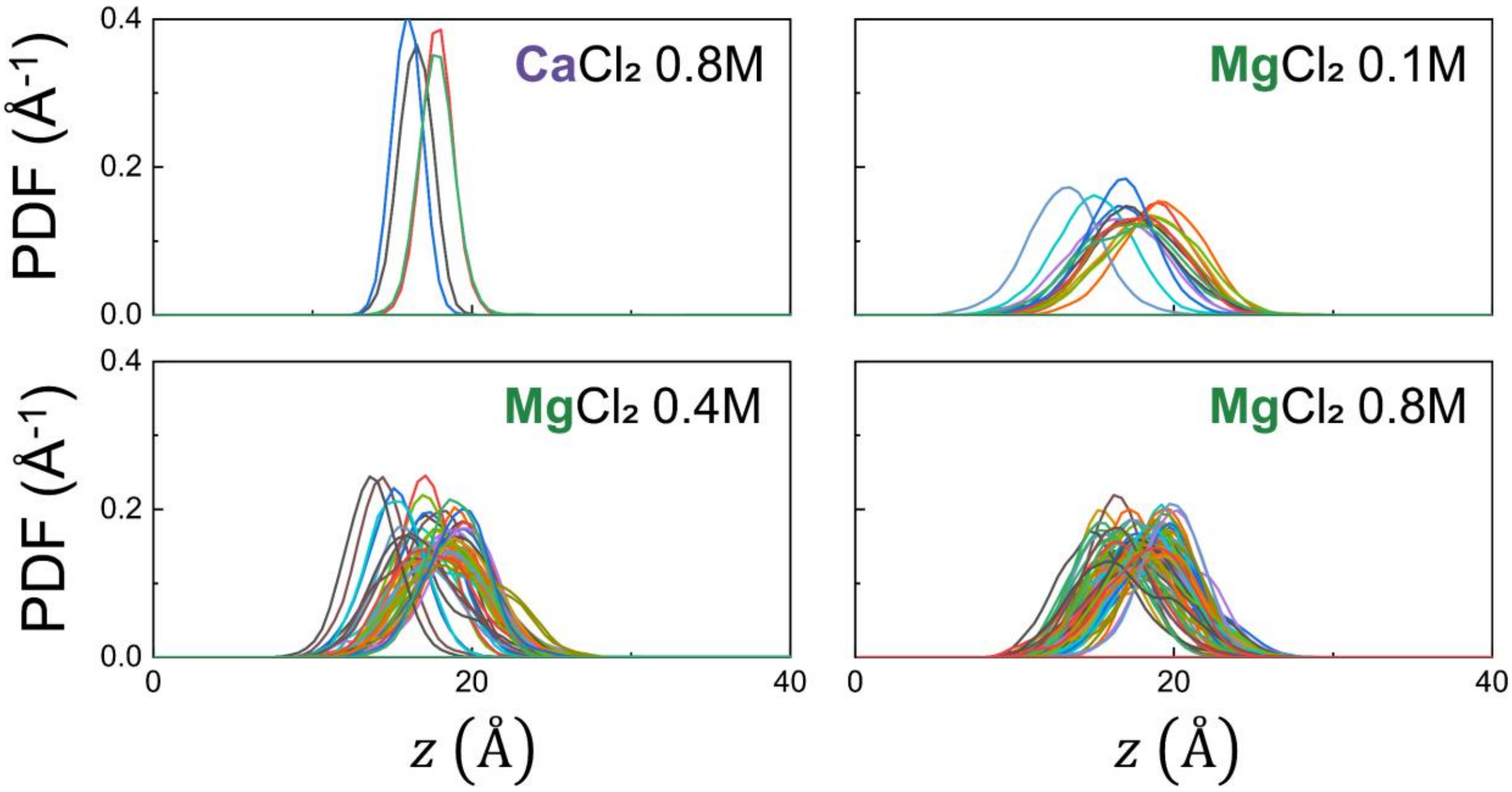

**FIG 12.** Distribution of the center of mass of trapped water molecules along the *z*-axis.

Furthermore, the time-averaged distribution of these water molecules along the *z*-axis, perpendicular to the membrane surface, shows that they are localized near the membrane interface (see Fig. 12). These observations demonstrate that the water molecules exhibiting long-time subdiffusion are strongly trapped by cations adsorbed at the membrane interface.

Notably, the number of trapped water molecules, shown in Table I, is significantly larger in the $MgCl_2$ system than in the $CaCl_2$ system. This observation is consistent with the results reported by Yi Dong *et al.*[15], that $Mg^{2+}$ ions retain their hydration shell even when they are located very close to the membrane surface.

## Appendix E. ROTATIONAL DYNAMICS OF INTERFACIAL WATER MOLECULES

To address an anonymous reviewer's comment, we discussed the relevance between the present work and previous studies in Refs. 33 and 34, which investigated the rotational relaxation of interfacial water molecules for various lipid bilayers with different headgroups. In Refs. 33 and 34, the authors reported that the number of hydrogen bonds between water molecules depends on the type of headgroup and that the rotational dynamics of interfacial water molecules becomes slower as the number of inter-water hydrogen bond per water molecule increases. In this section, we investigated the dependence of the water rotational dynamics on the number of inter-water hydrogen bonds per water molecule at various layers, systematically controlling salt types and concentrations. We found that, in our system, interaction between cations and water molecules is the dominant factor that retards rotational dynamics of water molecules and also decreases the number of inter-water hydrogen bonds per water molecule. For this reason, water rotational dynamics is significantly retarded, and the number of inter-water hydrogen bonds is diminished in layers 1 and 2 where cations are highly concentrated.

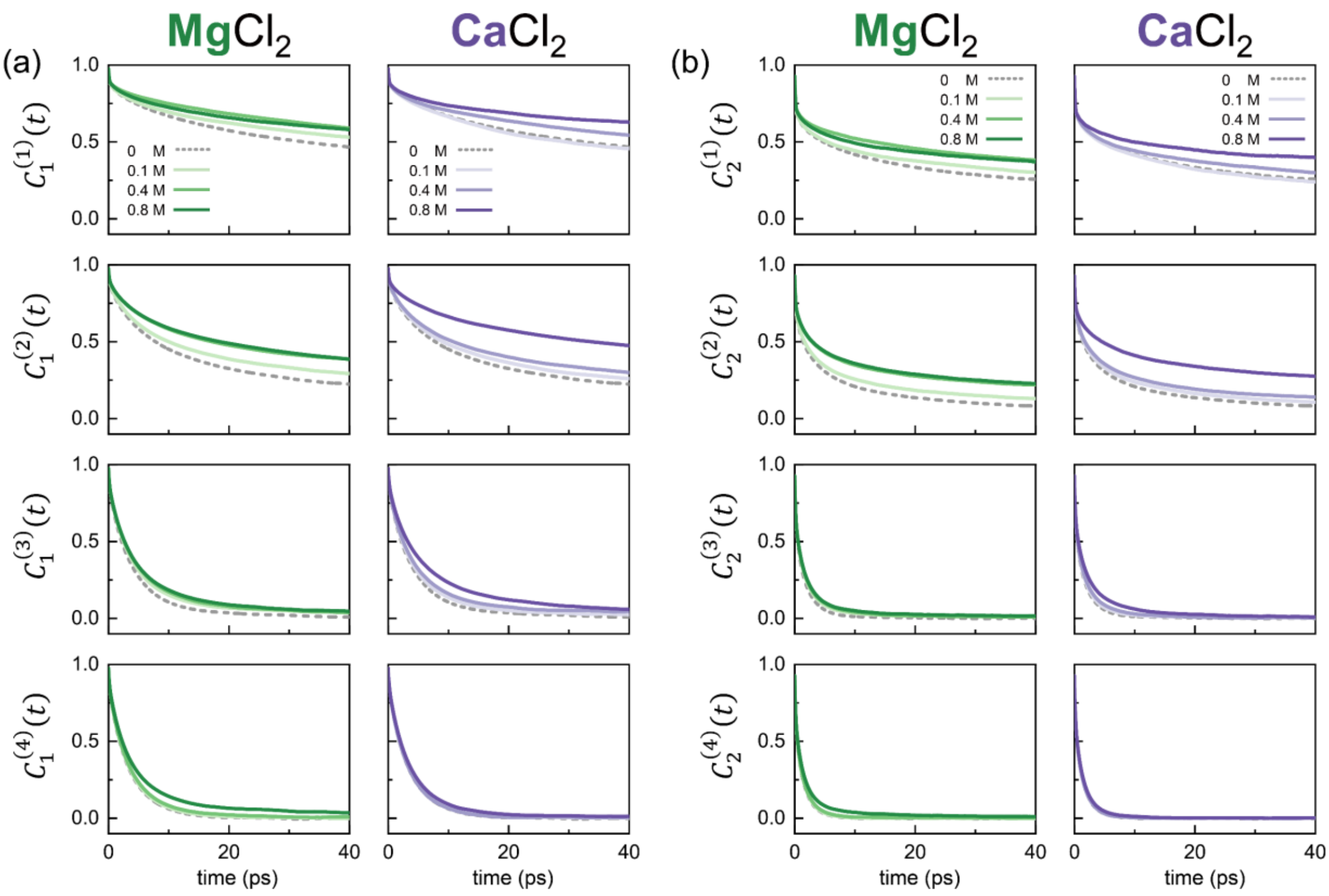

**FIG 13.** Effects of salt type and concentration on rotational dynamics of water molecules at various layers in the intermembrane space. (a) First-rank orientational correlation function, $C_1^{(n)}(t)$, of layer $n$. (b) Second-rank orientational correlation function, $C_2^{(n)}(t)$, of layer $n$.

To characterize rotational dynamics of interfacial water molecules, we investigated the first- and second-rank orientational correlation functions for each layer (see Fig. 13), defined by $C_1^{(n)}(t) \equiv \langle P_1(\cos\varphi(t))\rangle_n = \langle \cos\varphi(t)\rangle_n$ and $C_2^{(n)}(t) \equiv \langle P_2(\cos\varphi(t))\rangle_n = \langle (3\cos^2\varphi(t) - 1)/2\rangle_n$, with $\langle \ldots \rangle_n$ denoting the average taken over water molecules that remain within layer $n$ during the observation period. $\varphi(t)$ denotes the angle between the dipole moment vector of a water molecule at time $t$ and its initial dipole moment vector. Both $C_1^{(n)}(t)$ and $C_2^{(n)}(t)$ exhibit slower relaxation at higher salt concentrations for both $CaCl_2$ and $MgCl_2$ (Fig. 13). In the salt-free case, the rotational relaxation of water molecules in layers 1 and 2 is significantly slower than that in layers 3 and 4 due to water interactions with lipid headgroups, which are located across layers 1 and 2 [see Figs. 7(e) and 7(f)][33]. The salt-induced slowdown is also most pronounced in these layers. This results because cations concentrated in layers 1 and 2 effectively suppress rotational relaxation of water molecules in these layers [see Fig. 7(g)].

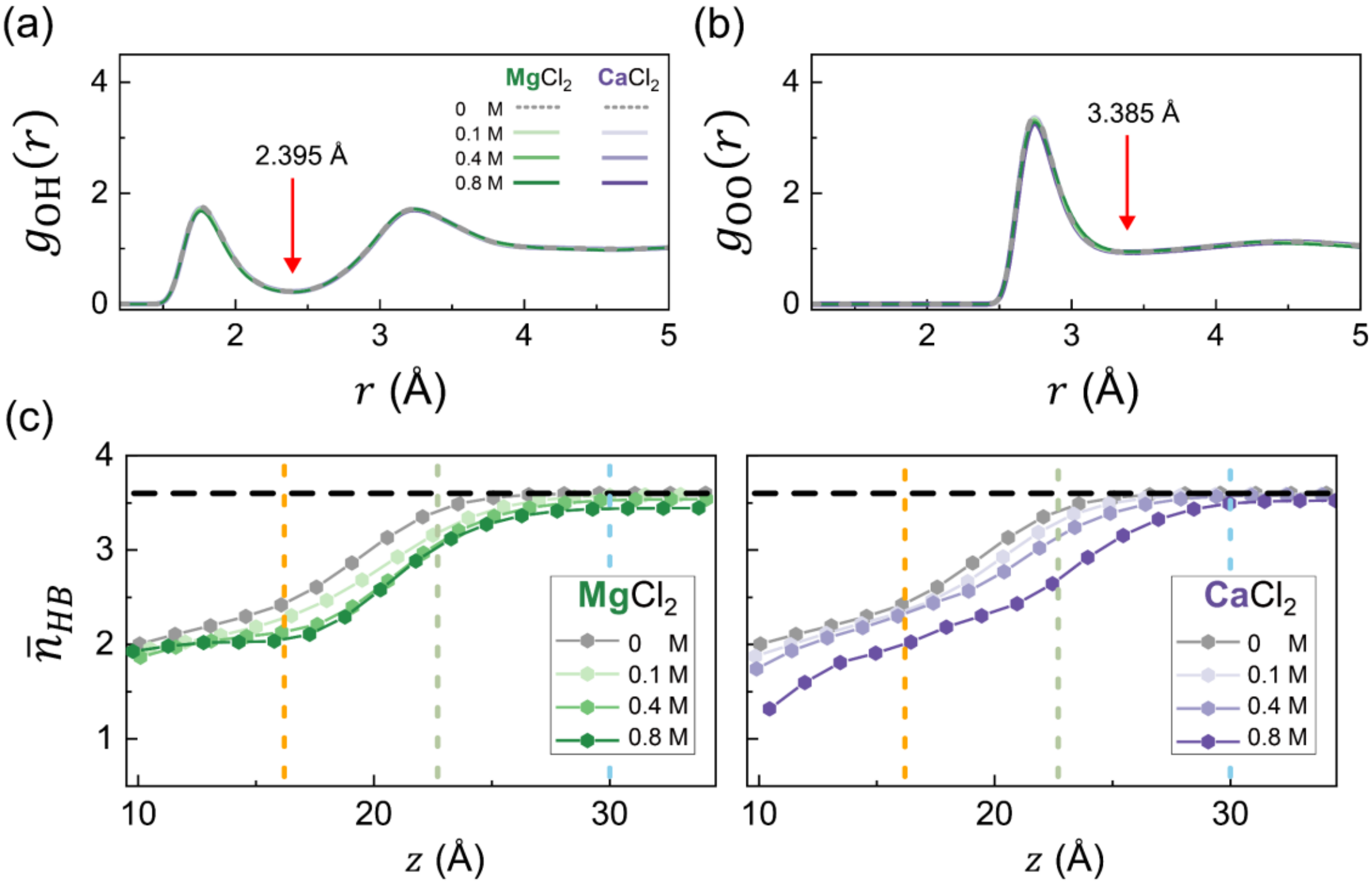

**FIG 14.** (a)-(b) Radial distribution functions (RDFs) between water molecules in $MgCl_2$ and $CaCl_2$ solutions: (a)

OH pair and (b) OO pair. In each panel, the arrow indicates the minimum position between the first and second RDF peaks. (c) $z$-dependent profile of the average hydrogen bond number, $\bar{\boldsymbol{n}}_{\boldsymbol{HB}}$, per water molecule. The horizontal black dashed line indicates the value, 3.6, in pure bulk water[64]. The vertical dashed lines indicate the layer boundaries, discussed in Figs. 4 and 7, located at z = 16.2, 22.7, and 30 Å.

To further understand the salt-dependent slowdown, we investigated the $z$-dependent profile of the average hydrogen bond number per water molecule. Hydrogen bonds were defined following Lee *et al.*[64]: the distance between O and H atoms is smaller than 2.395 Å, the distance between two O atoms is smaller than 3.385 Å [Figs. 14(a) and 14(b)], and the angle between the OH chemical bond and the hydrogen bond larger than 120°. Using this definition, we calculated the average hydrogen bond number, $\bar{n}_{HB}$, along the $z$-axis [Fig. 14(c)]. The results showed that $\bar{n}_{HB}$ decreases from its bulk value, 3.6[64], as $z$ decreases. Moreover, increasing the salt concentration leads to a further decrease in $\bar{n}_{HB}$, with a stronger decrease observed for $CaCl_2$ than for $MgCl_2$ at a given $z$.

These trends can be explained by the spatial distributions of lipid headgroups and ions reducing $\bar{n}_{HB}$ through their interactions with water molecules. Lipid headgroups, particularly phosphate groups, are mainly distributed across layers 1 and 2, and cations are likewise concentrated in this region [see Figs. 7(e) and 7(g)]. Due to their smaller hydration shell, $Ca^{2+}$ ions are predominantly localized in layers 1 and 2, rather than layers 3 and 4, whereas $Mg^{2+}$ ions with their larger hydration shell increasingly populate layers 3 and 4 as the salt concentration increases after adsorption sites become saturated. This leads to a reduction of $\bar{n}_{HB}$ in layer 4 for $MgCl_2$, unlike for $CaCl_2$. The results in Figs. 13 and 14 indicate that the slowdown of the rotational relaxation is primarily governed by cation-water interactions, rather than by water-water hydrogen bonding. This explains the apparent reversal of the trend reported in previous studies[33,34]. Although a more detailed investigation of this topic would be valuable, the present work primarily focuses on the transport dynamics of water molecules confined between lipid membranes, and we therefore leave this for future work.

## Appendix F. SALT-INDUCED WATER SQUEEZE-OUT EFFECT FOR CHOLESTEROL-INCORPORATED DMPC BILAYER MEMBRANE SYSTEM

Mixed membrane systems containing various lipid types[34,68,69] and cholesterol[70–74] have attracted considerable attention in recent studies. To examine whether the salt-induced water squeeze-out effect observed for DMPC bilayer membranes persists in a mixed membrane environment, we have performed additional *NpT* simulations for water molecules confined between DMPC/cholesterol mixed bilayers and calculated the area per lipid of the mixed membrane (see Section III of the main text for simulation details).

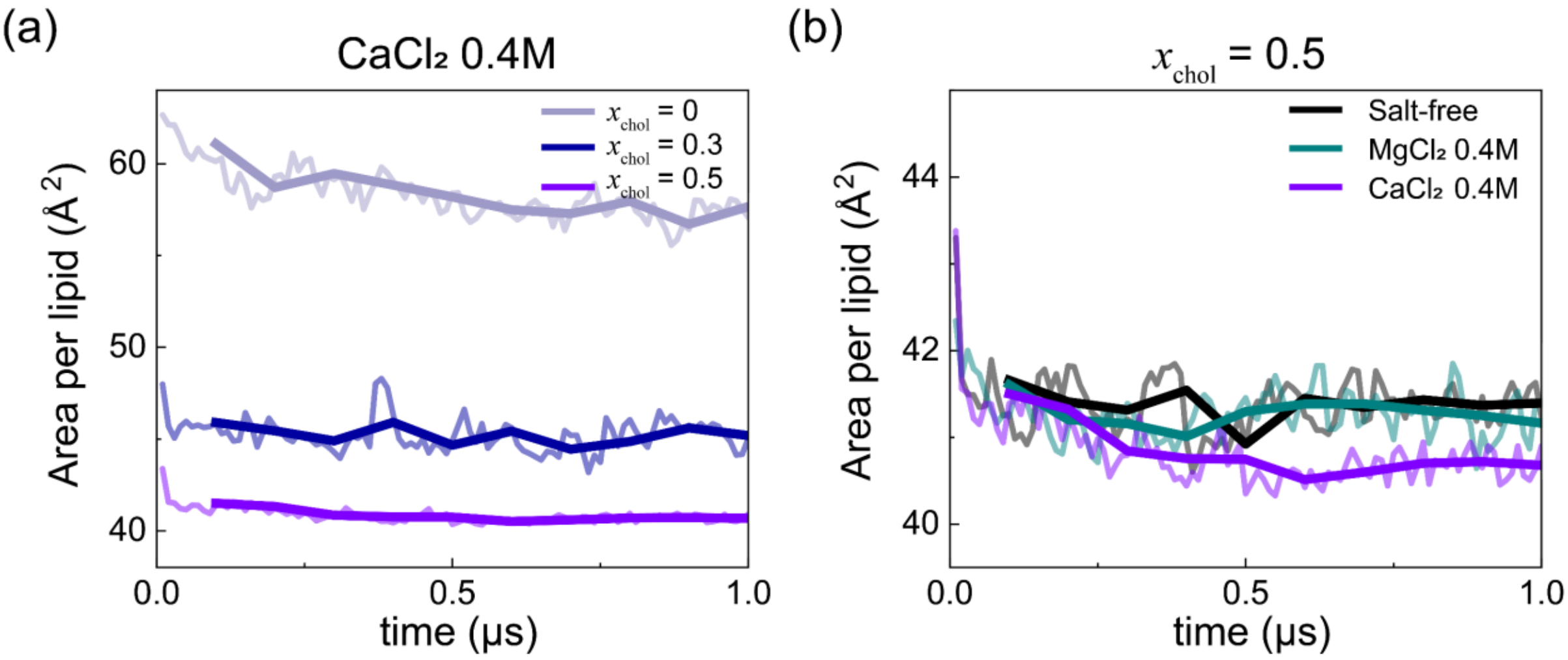

**FIG 15**. Time evolution of the area per lipid of DMPC/cholesterol mixed bilayer system. (thin lines) Block averages of the raw data, recorded every 10 ps, over 10 ns period (thick lines) Block averages of the raw data over 100 ns period. (a) The cholesterol mole fraction, $x_{\mathrm{chol}}$, dependence of the area per lipid at $CaCl_2$ concentration of 0.4M. (b) Salt-species dependence of the area per lipid at the cholesterol mole fraction of $x_{\mathrm{chol}} = 0.5$.

The incorporation of cholesterol into DMPC bilayers leads to a pronounced decrease in the area per lipid [Fig. 15(a)]. This result is consistent with previous reports that cholesterol alone induces the water squeeze-out effect in lipid membranes[71–73]. As shown in Fig. 15(b), the addition of 0.4M $CaCl_2$ results in a further reduction in the area per lipid, indicating that $Ca^{2+}$ ions induce an additional squeeze-out effect in the membranes already condensed by cholesterols. However, the $Ca^{2+}$-induced further reduction in the area per lipid shown in Fig. 15(b) for the cholesterol/DMPC bilayer system is marginal compared to the cholesterol-induced reduction in the area per lipid shown in Fig. 15(a) and the $Ca^{2+}$-induced reduction in the area per lipid shown in Fig. 2 for the DMPC bilayer system. In comparison, the addition of 0.4 M $MgCl_2$ does not induce a significant reduction in the area per lipid either for cholesterol/DMPC mixture bilayer system or for pure DMPC bilayer system [Figs. 15(b) and 2].

Finally, we note that lipid type, mixed bilayer composition, and temperature may also influence the area per lipid of the mixed membrane systems. A systematic investigation of these factors and their interplay in modulating the squeeze-out effects is left for future research.

# References

[1] H.Y. Ali Moussa, K.C. Shin, J. Ponraj, S.J. Kim, J.K. Ryu, S. Mansour, and Y. Park, "Requirement of Cholesterol for Calcium-Dependent Vesicle Fusion by Strengthening Synaptotagmin-1-Induced Membrane Bending," Advanced Science **10**(15), (2023).

[2] C. Allolio, and D. Harries, "Calcium Ions Promote Membrane Fusion by Forming Negative-Curvature Inducing Clusters on Specific Anionic Lipids," ACS Nano **15**(8), 12880–12887 (2021).

[3] B. Frankenhaeuser, and A.L. Hodgkin, "The action of calcium on the electrical properties of squid axons," J. Physiol. **137**(2), 218–244 (1957).

[4] O.H. Petersen, M. Michalak, and A. Verkhratsky, "Calcium signalling: Past, present and future," Cell Calcium **38**(3–4), 161–169 (2005).

[5] M.J. Berridge, M.D. Bootman, and H.L. Roderick, "Calcium signalling: dynamics, homeostasis and remodelling," Nat. Rev. Mol. Cell Biol. **4**(7), 517–529 (2003).

[6] S. EBASHI, "Third Component Participating in the Super precipitation of 'Natural Actomyosin,'" Nature **200**(4910), 1010–1010 (1963).

[7] E.W. Davie, and O.D. Ratnoff, "Waterfall Sequence for Intrinsic Blood Clotting," Science (1979). **145**(3638), 1310–1312 (1964).

[8] A. Romani, and A. Scarpa, "Regulation of cell magnesium," Arch. Biochem. Biophys. **298**(1), 1–12 (1992).

[9] C.F.E. Schroer, L. Baldauf, L. van Buren, T.A. Wassenaar, M.N. Melo, G.H. Koenderink, and S.J. Marrink, "Charge-dependent interactions of monomeric and filamentous actin with lipid bilayers," Proc. Natl. Acad. Sci. U. S. A. **117**(11), 5861–5872 (2020).

[10] N.L. Liebe, I. Mey, L. Vuong, F. Shikho, B. Geil, A. Janshoff, and C. Steinem, "Bioinspired Membrane Interfaces: Controlling Actomyosin Architecture and Contractility," ACS Appl. Mater. Interfaces **15**(9), 11586–11598 (2023).

[11] G.L. Eichhorn, and Y.A. Shin, "Interaction of metal ions with polynucleotides and related compounds. XII. The relative effect of various metal ions on DNA helicity," J. Am. Chem. Soc. **90**(26), 7323–7328 (1968).

[12] G.S. Manning, "The molecular theory of polyelectrolyte solutions with applications to the electrostatic properties of polynucleotides," Q. Rev. Biophys. **11**(2), 179–246 (1978).

[13] J.P. Abrahams, A.G.W. Leslie, R. Lutter, and J.E. Walker, "Structure at 2.8 Â resolution of F1-ATPase from bovine heart mitochondria," Nature **370**(6491), 621–628 (1994).

[14] E.M. Adams, C.B. Casper, and H.C. Allen, "Effect of cation enrichment on dipalmitoylphosphatidylcholine (DPPC) monolayers at the air-water interface," J. Colloid Interface Sci. **478**, 353–364 (2016).

[15] Y. Dong, L. Fu, J. Song, S. Zhang, X. Li, W. Fang, Q. Cui, and L. Gao, "Thermodynamic Driving Forces for Divalent Cations Binding to Zwitterionic Phospholipid Membranes," J. Phys. Chem. Lett. **13**(48), 11237–11244 (2022).

[16] R.J. Alsop, R. Maria Schober, and M.C. Rheinstädter, "Swelling of phospholipid membranes by divalent metal ions depends on the location of the ions in the bilayers," Soft Matter **12**(32), 6737–6748 (2016).

[17] S. Mallick, "Ion–Lipid Interactions in Biological Membranes: Insights from Combined Molecular Dynamics and Quantum Chemical Calculations," J. Phys. Chem. B, (2025).

[18] P. Alam, P. Kumar, H. Sahu, D. Sardana, P. Kundu, A.K. Chand, and S. Sen, "Effect of Divalent Cations on Polarity and Hydration at the Lipid/Water Interface Probed by 4-Aminophthalimide-Based Dyes," J. Phys. Chem. B **129**(3), 930–941 (2025).

[19] J. Yang, C. Calero, M. Bonomi, and J. Martí, "Specific Ion Binding at Phospholipid Membrane Surfaces," J. Chem. Theory Comput. **11**(9), 4495–4499 (2015).

[20] T. Shibata, *Pulse NMR Study of the Interaction of Calcium Ion with Dipalmitoylphosphatidylcholine Lamellae* (1990).

[21] C.Y. Tang, Z. Huang, and H.C. Allen, "Interfacial water structure and effects of Mg2+ and Ca 2+ binding to the COOH headgroup of a palmitic acid monolayer studied by sum frequency spectroscopy," Journal of Physical Chemistry B **115**(1), 34–40 (2011).

[22] W. Hua, D. Verreault, and H.C. Allen, "Solvation of Calcium-Phosphate Headgroup Complexes at the DPPC/Aqueous Interface," ChemPhysChem **16**(18), 3910–3915 (2015).

[23] P.M. Wiggins, "Role of water in some biological processes," Microbiol. Rev. **54**(4), 432–449 (1990).

[24] P. Ball, "Water is an active matrix of life for cell and molecular biology," Proceedings of the National Academy of Sciences **114**(51), 13327–13335 (2017).

[25] H. Seto, and T. Yamada, "Quasi-elastic neutron scattering study of the effects of metal cations on the hydration water between phospholipid bilayers," Appl. Phys. Lett. **116**(13), (2020).

[26] C. Calero, H. Stanley, and G. Franzese, "Structural Interpretation of the Large Slowdown of Water Dynamics at Stacked Phospholipid Membranes for Decreasing Hydration Level: All-Atom Molecular Dynamics," Materials **9**(5), 319 (2016).

[27] C.-J. Högberg, and A.P. Lyubartsev, "A Molecular Dynamics Investigation of the Influence of Hydration and Temperature on Structural and Dynamical Properties of a Dimyristoylphosphatidylcholine Bilayer," J. Phys. Chem. B **110**(29), 14326–14336 (2006).

[28] Y. Kazoe, K. Mawatari, L. Li, H. Emon, N. Miyawaki, H. Chinen, K. Morikawa, A. Yoshizaki, P.S. Dittrich, and T. Kitamori, "Lipid Bilayer-Modified Nanofluidic Channels of Sizes with Hundreds of Nanometers for Characterization of Confined Water and Molecular/Ion Transport," J. Phys. Chem. Lett. **11**(14), 5756–5762 (2020).

[29] J.B. Klauda, N. Kučerka, B.R. Brooks, R.W. Pastor, and J.F. Nagle, "Simulation-based methods for interpreting x-ray data from lipid bilayers," Biophys. J. **90**(8), 2796–2807 (2006).

[30] E. Zunzunegui-Bru, S.R. Alfarano, P. Zueblin, H. Vondracek, F. Piccirilli, L. Vaccari, S. Assenza, and R. Mezzenga, "Universality in the Structure and Dynamics of Water under Lipidic Mesophase Soft Nanoconfinement," ACS Nano **18**(32), 21376–21387 (2024).

[31] F. Martelli, J. Crain, and G. Franzese, "Network Topology in Water Nanoconfined between Phospholipid Membranes," ACS Nano **14**(7), 8616–8623 (2020).

[32] D. Sil, E. Osmanbasic, S.C. Mandal, A. Acharya, and C. Dutta, "Variable Non-Gaussian Transport of Nanoplastic on Supported Lipid Bilayers in Saline Conditions," J. Phys. Chem. Lett., 5428–5435 (2024).

[33] Y. Higuchi, Y. Asano, T. Kuwahara, and M. Hishida, "Rotational Dynamics of Water at the Phospholipid Bilayer Depending on the Head Groups Studied by Molecular Dynamics Simulations.," Langmuir **37**(17), 5329–5338 (2021).

[34] Md.K. Rahman, T. Yamada, N.L. Yamada, Y. Higuchi, and H. Seto, "Hydration Water Dynamics in Zwitterionic Phospholipid Membranes Mixed with Charged Phospholipids," J. Phys. Chem. B, (2025).

[35] J.A. Virtanen, K.H. Cheng, and P. Somerharju, "Phospholipid composition of the mammalian red cell membrane can be rationalized by a superlattice model," Proceedings of the National Academy of Sciences **95**(9), 4964–4969 (1998).

[36] F.Y. Hansen, G.H. Peters, H. Taub, and A. Miskowiec, "Diffusion of water and selected atoms in DMPC lipid bilayer membranes," J. Chem. Phys. **137**(20), 204910 (2012).

[37] J. Jang, S. Kim, and K. Eom, "NaCl increases the dielectric constant of nanoconfined water in phospholipid multilamellar vesicle by enhancing intermolecular orientation correlation rather than rotational freedom of individual molecules," Chem. Phys. Lett. **780**, 138912 (2021).

[38] L.J. Lis, W.T. Lis, V.A. Parsegian, and R.P. Rand, "Adsorption of divalent cations to a variety of phosphatidylcholine bilayers," Biochemistry **20**(7), 1771–1777 (1981).

[39] K. Åman, E. Lindahl, O. Edholm, P. Håkansson, and P.-O. Westlund, "Structure and Dynamics of Interfacial Water in an Lα Phase Lipid Bilayer from Molecular Dynamics Simulations," Biophys. J. **84**(1), 102–115 (2003).

[40] S. Mallick, and N. Agmon, "Lateral diffusion of ions near membrane surface," Physical Chemistry Chemical Physics **26**(28), 19433–19449 (2024).

[41] S. Song, S.J. Park, M. Kim, J.S. Kim, B.J. Sung, S. Lee, J.-H. Kim, and J. Sung, "Transport dynamics of complex fluids," Proceedings of the National Academy of Sciences **116**(26), 12733–12742 (2019).

[42] M. Lee, E. Lee, J.-H. Kim, H. Hwang, M. Cho, and J. Sung, "Transport Dynamics of Water Molecules Confined between Lipid Membranes," J. Phys. Chem. Lett. **15**(16), 4437–4443 (2024).

[43] A. Rahman, K.S. Singwi, and A. Sjölander, "Theory of Slow Neutron Scattering by Liquids. I," Physical Review **126**(3), 986–996 (1962).

[44] A. Rahman, "Correlations in the Motion of Atoms in Liquid Argon," Physical Review **136**(2A), A405–A411 (1964).

[45] R. Zhang, T.A. Cross, X. Peng, and R. Fu, "Surprising Rigidity of Functionally Important Water Molecules Buried in the Lipid Headgroup Region," J. Am. Chem. Soc. **144**(17), 7881–7888 (2022).

[46] E. Yamamoto, T. Akimoto, Y. Hirano, M. Yasui, and K. Yasuoka, "Power-law trapping of water molecules on the lipid-membrane surface induces water retardation.," Phys. Rev. E Stat. Nonlin. Soft Matter Phys. **87**(5), 052715 (2013).

[47] S. Jo, J.B. Lim, J.B. Klauda, and W. Im, "CHARMM-GUI Membrane Builder for mixed bilayers and its application to yeast membranes.," Biophys. J. **97**(1), 50–8 (2009).

[48] J. Lee, X. Cheng, J.M. Swails, M.S. Yeom, P.K. Eastman, J.A. Lemkul, S. Wei, J. Buckner, J.C. Jeong, Y. Qi, S. Jo, V.S. Pande, D.A. Case, C.L. Brooks, A.D. MacKerell, J.B. Klauda, and W. Im, "CHARMM-GUI Input Generator for NAMD, GROMACS, AMBER, OpenMM, and CHARMM/OpenMM Simulations Using the CHARMM36 Additive Force Field," J. Chem. Theory Comput. **12**(1), 405–413 (2016).

[49] E.L. Wu, X. Cheng, S. Jo, H. Rui, K.C. Song, E.M. Dávila-Contreras, Y. Qi, J. Lee, V. Monje-Galvan, R.M. Venable, J.B. Klauda, and W. Im, "CHARMM-GUI *Membrane Builder* toward realistic biological membrane simulations," J. Comput. Chem. **35**(27), 1997–2004 (2014).

[50] J. Lee, M. Hitzenberger, M. Rieger, N.R. Kern, M. Zacharias, and W. Im, "CHARMM-GUI supports the Amber force fields," J. Chem. Phys. **153**(3), (2020).

[51] C.J. Dickson, R.C. Walker, and I.R. Gould, "Lipid21: Complex Lipid Membrane Simulations with AMBER," J. Chem. Theory Comput. **18**(3), 1726–1736 (2022).

[52] P. Li, L.F. Song, and K.M. Merz, "Systematic parameterization of monovalent ions employing the nonbonded model," J. Chem. Theory Comput. **11**(4), 1645–1657 (2015).

[53] Z. Li, L.F. Song, P. Li, and K.M. Merz, "Systematic Parametrization of Divalent Metal Ions for the OPC3, OPC, TIP3P-FB, and TIP4P-FB Water Models," J. Chem. Theory Comput. **16**(7), 4429–4442 (2020).

[54] A. Sengupta, Z. Li, L.F. Song, P. Li, and K.M. Merz, "Parameterization of Monovalent Ions for the OPC3, OPC, TIP3P-FB, and TIP4P-FB Water Models," J. Chem. Inf. Model. **61**(2), 869–880 (2021).

[55] P. Li, and K.M. Merz, "Taking into account the ion-induced dipole interaction in the nonbonded model of ions," J. Chem. Theory Comput. **10**(1), 289–297 (2014).

[56] T. Darden, D. York, and L. Pedersen, "Particle mesh Ewald: An N·log(N) method for Ewald sums in large systems," J. Chem. Phys. **98**(12), 10089–10092 (1993).

[57] Y. Von Hansen, S. Gekle, and R.R. Netz, "Anomalous anisotropic diffusion dynamics of hydration water at lipid membranes," Phys. Rev. Lett. **111**(11), 1–8 (2013).

[58] S. Deublein, S. Reiser, J. Vrabec, and H. Hasse, "A set of molecular models for alkaline-earth cations in aqueous solution," Journal of Physical Chemistry B **116**(18), 5448–5457 (2012).

[59] S. Roy, S.M. Gruenbaum, and J.L. Skinner, "Theoretical vibrational sum-frequency generation spectroscopy of water near lipid and surfactant monolayer interfaces," J. Chem. Phys. **141**(18), (2014).

[60] M. Doğangün, P.E. Ohno, D. Liang, A.C. McGeachy, A.G. Bé, N. Dalchand, T. Li, Q. Cui, and F.M. Geiger, "Hydrogen-Bond Networks near Supported Lipid Bilayers from Vibrational Sum Frequency Generation Experiments and Atomistic Simulations," J. Phys. Chem. B **122**(18), 4870–4879 (2018).

[61] W. Zhao, D.E. Moilanen, E.E. Fenn, and M.D. Fayer, "Water at the Surfaces of Aligned Phospholipid Multibilayer Model Membranes Probed with Ultrafast Vibrational Spectroscopy," J. Am. Chem. Soc. **130**(42), 13927–13937 (2008).

[62] J.A. Mondal, S. Nihonyanagi, S. Yamaguchi, and T. Tahara, “Three Distinct Water Structures at a Zwitterionic Lipid/Water Interface Revealed by Heterodyne-Detected Vibrational Sum Frequency Generation,” J. Am. Chem. Soc. **134**(18), 7842–7850 (2012).

[63] L.B. Dreier, A. Wolde-Kidan, D.J. Bonthuis, R.R. Netz, E.H.G. Backus, and M. Bonn, “Unraveling the Origin of the Apparent Charge of Zwitterionic Lipid Layers.,” J. Phys. Chem. Lett. **10**(20), 6355–6359 (2019).

[64] E. Lee, A. Kundu, J. Jeon, and M. Cho, “Water hydrogen-bonding structure and dynamics near lipid multibilayer surface: Molecular dynamics simulation study with direct experimental comparison,” J. Chem. Phys. **151**(11), 114705 (2019).

[65] M.C. Gurau, G. Kim, S.M. Lim, F. Albertorio, H.C. Fleisher, and P.S. Cremer, “Organization of Water Layers at Hydrophilic Interfaces,” ChemPhysChem **4**(11), 1231–1233 (2003).

[66] B. Das, and A. Chandra, “Ab Initio Molecular Dynamics Study of Aqueous Solutions of Magnesium and Calcium Nitrates: Hydration Shell Structure, Dynamics and Vibrational Echo Spectroscopy,” Journal of Physical Chemistry B **126**(2), 528–544 (2022).

[67] S. Jo, T. Kim, V.G. Iyer, and W. Im, “CHARMM-GUI: A web-based graphical user interface for CHARMM,” J. Comput. Chem. **29**(11), 1859–1865 (2008).

[68] A. Majumder, Y. Gu, Y.C. Chen, X. An, B.M. Reinhard, and J.E. Straub, “Probing the Origins of the Disorder-to-Order Transition of a Modified Cholesterol in Ternary Lipid Bilayers,” J. Am. Chem. Soc. **146**(40), 27725–27735 (2024).

[69] G. Shahane, W. Ding, M. Palaiokostas, and M. Orsi, “Physical properties of model biological lipid bilayers: insights from all-atom molecular dynamics simulations,” J. Mol. Model. **25**(3), 76 (2019).

[70] H. Orlikowska-Rzeznik, J. Versluis, H.J. Bakker, and L. Piatkowski, “Cholesterol Changes Interfacial Water Alignment in Model Cell Membranes,” J. Am. Chem. Soc. **146**(19), 13151–13162 (2024).

[71] M.D. Elola, and J. Rodriguez, “Influence of Cholesterol on the Dynamics of Hydration in Phospholipid Bilayers,” J. Phys. Chem. B **122**(22), 5897–5907 (2018).

[72] B.B. Issack, and G.H. Peslherbe, “Effects of Cholesterol on the Thermodynamics and Kinetics of Passive Transport of Water through Lipid Membranes,” J. Phys. Chem. B **119**(29), 9391–9400 (2015).

[73] E. Falck, M. Patra, M. Karttunen, M.T. Hyvönen, and I. Vattulainen, “Lessons of Slicing Membranes: Interplay of Packing, Free Area, and Lateral Diffusion in Phospholipid/Cholesterol Bilayers,” Biophys. J. **87**(2), 1076–1091 (2004).

[74] K. Shikata, K. Kasahara, N.M. Watanabe, H. Umakoshi, K. Kim, and N. Matubayasi, "Dual effect of cholesterol on interfacial water dynamics in lipid membranes: Interplay between membrane packing and hydration," J. Chem. Phys. **163**(24), (2025).